\documentstyle[12pt]{article}
\renewcommand{\baselinestretch}{1.0}
   \newcounter{multieqn}
\renewcommand{\theequation}{\arabic{section}.\arabic{equation}}
\begin{document}
%
\setcounter{footnote}{3}
\title{Theory of quantum radiation observed as sonoluminescence}
\author{Claudia Eberlein%
\thanks{e-mail to: claudia@cromwell.physics.uiuc.edu}\\
	Department of Physics\\
	University of Illinois at Urbana-Champaign\\
	Urbana, IL 61801-3080\\
        U.S.A.}
\date{(Submitted June 7, 1995)}
\maketitle
\begin{abstract}
	Sonoluminescence is explained in terms of quantum radiation by
moving interfaces between media of different polarizability. In a
stationary dielectric the zero-point fluctuations of the
electromagnetic field excite virtual two-photon states which become
real under perturbation due to motion of the dielectric. The
sonoluminescent bubble is modelled as an optically empty cavity in a
homogeneous dielectric. The problem of the photon emission by a cavity
of time-dependent radius is handled in a Hamiltonian formalism which
is dealt with perturbatively up to first order in the velocity of the
bubble surface over the speed of light. A parameter-dependence of the
zero-order Hamiltonian in addition to the first-order perturbation
calls for a new perturbative method combining standard perturbation
theory with an adiabatic approximation. In this way the transition
amplitude from the vacuum into a two-photon state is obtained, and
expressions for the single-photon spectrum and the total energy
radiated during one flash are given both in full and in the
short-wavelengths approximation when the bubble is larger than the
wavelengths of the emitted light. A model profile is assumed for the
time-dependence of the bubble during the collapse, and in this model
the radiated energy and the spectrum are calculated numerically and in
the short-wavelengths limit also analytically. It is shown
analytically that the spectral density has the same
frequency-dependence as black-body radiation; this is purely an effect
of correlated quantum fluctuations at zero temperature. The present
theory clarifies a number of hitherto unsolved problems and suggests
explanations for several more. Possible experiments that discriminate
this from other theories of sonoluminescence are proposed.

\vspace*{5mm}
PACS numbers: 03.70.+k, 11.10.-z, 42.50.Lc, 78.60.Mq
\end{abstract}
\setcounter{footnote}{0}

\section{INTRODUCTION}
\setcounter{equation}{0}
\subsection{State of the art}

Sonoluminescence is the phenomenon of light emission by sound-driven
gas bubbles in fluids, ordinarily air bubbles in water. Sound makes
bubbles collapse or expand, and a rapid flash of light is observed
after each collapse. This phenomenon has been known for 60 years
\cite{first}, but came under systematic investigation only recently
when experimentalists succeeded in trapping bubbles and maintaining
sonoluminescence as a stable process over hours or even days
\cite{stable1,stable2}.

During stable sonoluminescence \cite{stable1,stable2} a bubble is
trapped at the pressure antinode of a standing sound wave, which
typically has a frequency of about 25 kHz. With an astonishing
clocklike precision the bubble sends off one sharp flash of light per
acoustic cycle. Less than 10 ps is commonly given as a conservative
estimate of the pulse length. The observed jitter has been found to be
extremely small and to show curious phase properties whose origin
could so far not be identified \cite{chaos}. The spectral density of
the light emitted drops with wavelength and resembles the tail of a
black-body spectrum of several tens of thousand Kelvin \cite{spec}.

Whereas the dynamics of the bubble motion has been successfully
explained and a theoretical model by L\"{o}fstedt, Barber, and
Putterman \cite{hydro} based on rather involved hydrodynamic
calculations reproduces the experimentally measured time-dependence of
the bubble radius \cite{miemeas} remarkably well, the process of the
light emission has so far defied any theoretical elucidation. That is
why the present paper focuses on the radiation process, making use of
the knowledge about the hydrodynamics of the bubble motion as input.

There have been several attempts of explaining the light seen in
sonoluminescence. The apparent similarity of the spectrum to a thermal
spectrum has led to the hypothesis that the light might come from a
process of black-body radiation or bremsstrahlung \cite{spec,brems}.
Along this line it has been argued that the gas in the collapsing
bubble is compressed so strongly that a plasma is formed which then
radiates. However, one can quickly convince oneself that neither
black-body radiation nor bremsstrahlung can possibly account for the
radiation observed in sonoluminescence. Either of them would lead to a
continuous spectrum whose major part would lie below the absorption edge of
water at 180nm and would therefore be absorbed by it. Estimating from
the corresponding visible part of the spectrum the amount of energy
that would be absorbed, one obtains such a large number that one would
expect to see rather obvious macroscopic consequences of the
absorption \cite{priv}, as for instance dissociation of the water
molecules, formation of radicals etc., which have not been observed.
Moreover, black-body radiation is an equilibrium phenomenon and
involves several atomic transitions; it could very unlikely explain
pulse lengths of less than 10ps. Neither is any explanation involving
bremsstrahlung satisfactory, because it would entail the presence of
free electrons and rather slow recombination radiation.

Rather more convincing is Suslick's theory \cite{susl} which explains
the sonoluminescence spectra on the basis of pressure-broadened
rotational and vibrational lines in diatomic emission spectra. For
silicone-oil sonoluminescence one finds an excellent agreement of
synthetic and observed spectra by considering emission from
excited-state $\rm C_{2}$ \cite{susl}. For water, however, any
attempts to model the spectrum on the basis of this theory have so far
been unsuccessful, although the well-known 310nm system of OH is
thought to be largely responsible for the broad peak around this
wavelength in the observed spectrum \cite{priv,fareast}.

The most recent speculation on the nature of sonoluminescence
radiation is a theory of collision-induced emission \cite{collis},
which, however, in its present version contains still too many
indeterminate points and adjustable parameters to permit a judgement
on its tenability.

\subsection{Quantum vacuum radiation as a candidate}

This article pursues a line of thought loosely inspired by Schwinger
\cite{jschw} who suggested that sonoluminescence could be some kind of
dynamic Casimir effect, which the present writer agrees with in so far
as the light emission observed in sonoluminescence has, just like the
Casimir effect, its origin in the interaction of the vacuum
fluctuations of the quantized electromagnetic field with a dielectric
medium. Sonoluminescence is, however, much more closely related to the
Unruh effect.

Let first the Casimir effect \cite{casi,milo} be recalled: two
parallel conducting or dielectric plates in vacuum feel an attractive
force which arises from the boundary conditions the plates impose on
the vacuum electromagnetic field. In a more intuitive picture one can
understand the Casimir effect in terms of van-der-Waals forces; the
electromagnetic zero-point fluctuations induce local fluctuating
dipoles in each of the plates and because of the spatial correlations
of the fluctuations the interaction of these dipoles leads to a net
attractive force.

The Unruh effect \cite{unruh,birdav} is a dynamic generalization of
the Casimir effect and predicts radiation by non-inertially moving
mirrors. This phenomenon is not exclusive to perfect mirrors, i.e.
perfect conductors; quantum radiation by moving dielectrics has also
been investigated \cite{GBCE}, and moreover some of the pathological
points of the perfect-reflector theories can be circumvented in the
more physical case of dielectrics. Again, the intuitive picture of the
process is that the zero-point electromagnetic field excites
fluctuating dipoles in the (perfect or imperfect) mirror and these
dipoles are the source of radiation when the mirror moves
non-uniformly.

A more rigorous way of understanding why a moving mirror that
interacts with the vacuum fluctuations of the quantized photon field
emits radiation, is to start by considering a nominally stationary
mirror, with the intention of eventually applying the
fluctuation-dissipation theorem. The radiation pressure on the mirror
is given by the vacuum expectation value of the force operator $\cal
F$, which is obtained from the stress-energy-momentum tensor of the
electromagnetic field subjected to appropriate boundary conditions on
the surface of the mirror. The net force on a single stationary mirror
in vacuum is of course zero by virtue of translation invariance,
$$
	F = \langle 0 | {\cal F} | 0 \rangle = 0 \;.
$$
However, the mean-square deviation of this force does not vanish,
since the force operator $\cal F$ does not commute with the
Hamiltonian. In other words, the mirror is exposed to
radiation-pressure fluctuations, whose mean-square deviation is given
by
$$
	\triangle F^{2} = \langle 0 | {\cal F}^{2} | 0 \rangle
                  - \langle 0 | {\cal F} | 0 \rangle^{2} \;.
$$
Knowing that the force operator $\cal F$ is (just like the
Hamiltonian) a functional that is quadratic in the field operators,
i.e. quadratic in the photon annihilation and creation operators, one
can use the decomposition of the identity into projection operators
onto a complete set of photon eigenstates, of which then only
two-photon states survive, and rewrite \cite{fn1}:
\begin{equation}
	\triangle F^{2} = \frac{1}{2}\: \int{\rm d}k \int{\rm d}k' \;
                          | \langle 0 | {\cal F} | k, k' \rangle |^{2} \;.
\label{statforce}
\end{equation}
This means that virtual two-photon states are perpetually excited by
the mirror in the vacuum, in accordance with the fluctuating radiation
pressure. Yet the fluctuating forces on the left and right sides of
the mirror are balanced against each other, so that no mean
radiation pressure acts on the mirror. By virtue of Lorentz
invariance, the same is true for a mirror that moves with constant
velocity.

However, when the mirror moves non-inertially, the radiation-pressure
fluctuations on opposite sites of the mirror are out of balance and
the mirror experiences a non-vanishing frictional force. The
virtual two-photon states turn into real states, and the loss of
momentum by the radiation of the photon pairs provides the physical
mechanism for the friction felt by the mirror. The
fluctuation-dissipation theorem puts this into formulae and
interrelates the power spectrum of the fluctuations on the stationary
mirror and the dissipative part of the response function that connects
the force on the moving mirror to its velocity \cite{CEfluc}.

It is a well-established fact that radiation by moving mirrors shows
thermal properties although one is dealing with {\em zero}-temperature
quantum field theory. The original statement of the Unruh effect
\cite{unruh,birdav} is that a mirror moving with constant proper
acceleration $a$ in vacuum appears to be radiating particles as if it
were a black body at a temperature $T_{\rm Unruh}=\hbar a /(2\pi
k_{\rm B} c)$. The reason for this behaviour is that the photons are
radiated in correlated pairs, in the language of quantum optics ---
they form a two-mode squeezed state, and the observation of the
single-photon spectrum involves a tracing over the other photon of the
pair which is well-known to entail thermal properties of the state
\cite{knight}.  Formally this connection is established by
representing the two-mode state in a dual Hilbert space and making
contact with the theory of thermofield dynamics \cite{umez}.

As to an experimental verification of the Unruh effect, the record is
empty. Understandably so, because the Unruh temperature is tiny for
commonly achievable accelerations. The only viable suggestion for
an experiment has come from Yablonovitch \cite{yabl}, who thought that
the sudden ionization of a gas or a semiconductor crystal might
produce an accelerating discontinuity in the refractive index fast
enough to radiate a measurable amount of photons.

{}From all of the above, quantum vacuum radiation seems to be a good
candidate for explaining the radiation process in sonoluminescence.
The surface of the bubble is the moving interface of discontinuous
polarizability, i.e. the moving mirror. In the visible range water has
a refractive index of 1.3, and the gas inside the bubble has a
refractive index of practically 1 even when strongly compressed.
Although the discontinuity of 0.3 in the refractive index is not huge,
it is large enough to radiate an appreciable number of photons if the
motion is sufficiently fast. In fact, the discontinuity in the
refractive index will enter the final results for the radiated
spectrum merely as a prefactor, and hence only its order of magnitude
is important.

Of much greater significance is the highly non-linear dynamics of the
bubble motion. At the point when the bubble collapses and starts
re-expanding, the velocity of the interface changes its sign. As known
from experiments \cite{miemeas} as well as from model calculations of
the bubble dynamics \cite{hydro,brems} this turn-around is extremely
fast, which means that tremendous accelerations and higher moments of
the motion are involved. The present theory predicts a burst of
photons as a consequence.

As it will be shown in more detail in the course of this article, the
theory of quantum vacuum radiation resolves several to date
unexplained issues.

The fact that the photons are radiated in correlated pairs leads
to thermal properties of the one-photon spectrum, unrelated to the
temperature in the bubble which is presumably far too small to cause
any major effect.

In accordance with a weakly frequency-dependent refractive index the
radiation spectrum shows features at the resonance frequencies of the
dipolar molecules in the medium. As water molecules are highly
polarizable, this is to be expected a discernible effect; it explains
the relation of the peak around 310nm in the spectrum to the
well-known OH line.

Barely any photons are created below the absorption edge of water,
as the polarizability is far too small in this region. Therefore, few
photons are absorbed and no macroscopically noticeable changes of
the water are to be expected.

The pulse length predicted by the theory of quantum vacuum radiation is
of the order of the time it takes for the zero-point fluctuations to
correlate around the bubble. With bubble sizes of around $1\mu{\rm m}$
or less, the time light takes to cross the bubble is in the
femtosecond range. Otherwise the time-scale is of course influenced by
the dynamics of the motion of the bubble interface at and just after
the collapse.

\subsection{Outline and overview}

The theory of quantum vacuum radiation by a gas bubble in water will be
expounded in the following sections. Water will be understood as a
non-absorbing dielectric describable by a constant refractive index.
This is a good approximation in the spectral region of interest where
water is only weakly dispersive. By virtue of adiabaticity the
refractive index $n$ can be replaced by $n(\omega)$ in the end result
for the radiated spectrum. The gas inside the bubble is optically so
thin, even at the collapse of the bubble, that its refractive index
will be assumed to be 1 throughout the calculation.

The bubble will be considered as externally driven, i.e. the radius of
the bubble as a prescribed function of time; the hydrodynamics of the
bubble motion is not the concern of this paper. However, the
back-reaction of the radiation process on the motion of the bubble will
be specified.

Hence the problem is reduced to a model of a spherical cavity of
radius $R(t)$ in a homogeneous non-dispersive dielectric described by
a constant refractive index $n$. The radiated spectral density will be
obtained as a functional of $R(t)$.

The next section deals with the quantization of the photon field in
the presence of a stationary spherical bubble in a homogeneous
dielectric. In section III the Schr{\"o}dinger equation for the photon
state-vector is written down, and the vacuum-to-two-photon
transition-amplitude is calculated by a method of time-dependent
perturbation theory that accommodates both an adiabatic
time-dependence of the Hamiltonian and a perturbative addition to the
zero-order Hamiltonian \cite{thesis}. Section IV states and examines
the results for the radiated energy and the spectral density. The
appearance of a thermal-like spectrum is demonstrated and numerical
results are presented. Finally, section V gives a summary and a
critical reflection on the theory of quantum vacuum radiation by
sonoluminescent bubbles. Strengths and weaknesses of the theory are
scrutinized, and open questions are voiced.

Readers not interested in the technicalities of the theory are
encouraged to look over section IV for an aggregate of the essential
results and to read section V for a guided summary and interpretation
of all results.

Several appendices contribute technical details necessary for the
clarity of the presentation. Appendix A calculates the Hamiltonian for
a uniformly moving dielectric in preparation for section III. Appendix B
gives the mode-expansion for the Helmholtz equation in spherical
coordinates, which is essential throughout the paper. The force on a
stationary dielectric is determined in appendix C.

CGS units are used everywhere in the paper; $\hbar$ and $c$ are set
equal to 1 unless explicitly indicated. All special functions are
defined as in refs. \cite{GR,AS}.

\section{QUANTIZATION OF THE PHOTON FIELD}
\setcounter{equation}{0}
The Hamiltonian for the electromagnetic field in the presence of a
medium with dielectric function $\varepsilon({\bf r})$ reads
\begin{equation}
	H_0 = \frac{1}{2}\: \int {\rm d}^3 {\bf r}\: \left(
	\frac{{\bf D}^2}{\varepsilon} + {\bf B}^2 \right)\;.
\label{h0}
\end{equation}
A bubble of radius $R$ is described by
\begin{equation}
	\varepsilon(r;R) = 1 + (n^2 -1)\: \theta(r-R)\;,
\label{eps}
\end{equation}
where $\theta$ is the Heaviside step function. This is to say that the
dielectric constant equals 1 in the interior of the bubble and $n^2$
in the surrounding medium. The Maxwell equations imply continuity
conditions for the fields across the boundary; these are:
\renewcommand{\arraystretch}{1.2}
\begin{equation}
	\begin{array}{l}
	{\bf D}_{\perp}\ {\rm and}\
	\displaystyle{\frac{{\bf D}_{\parallel}}{\varepsilon}}\
	{\rm continuous\:,}\\
	{\bf B}\ {\rm continuous}\:,
	\end{array}
\label{contgen}
\end{equation}
or in spherical coordinates
\renewcommand{\arraystretch}{1.5}
\begin{equation}
	\begin{array}{l}
	{\bf D}_{r}^{\rm inside} = {\bf D}_{r}^{\rm outside}\:,\ \
	{\bf D}_{\theta}^{\rm inside} =
	\displaystyle{\frac{{\bf D}_{\theta}^{\rm outside}}{n^2}}\:,\ \
	{\bf D}_{\phi}^{\rm inside} =
	\displaystyle{\frac{{\bf D}_{\phi}^{\rm outside}}{n^2}}\:,\\
	{\bf B}_{r}^{\rm inside} = {\bf B}_{r}^{\rm outside}\:,\ \
	{\bf B}_{\theta}^{\rm inside} = {\bf B}_{\theta}^{\rm outside}\:,\ \
	{\bf B}_{\phi}^{\rm inside} = {\bf B}_{\phi}^{\rm outside}\:.
	\end{array}
\label{contexpl}
\end{equation}
\renewcommand{\arraystretch}{1}

The Hamiltonian (\ref{h0}) depends parametrically on the bubble radius
$R$ via the dielectric function $\varepsilon(r;R)$. Although a problem
with a varying bubble size is aspired to be solved, for the purpose of
quantizing the photon field the radius of the bubble will be kept
constant. In order to quantize the system for a time-dependent radius
$R(t)$ one would need to know the eigenfunctions of the time-dependent
Hamiltonian (\ref{ham}); but knowing them would amount to the exact
solution of the whole problem which is of course unachievable. As the
calculation to follow in the next section will employ perturbation
theory to first order in the velocity of the bubble surface
$\beta=\dot{R}(t)$ over the speed of light in vacuum, a quantization
for constant $R$ is fully sufficient.

The field is quantized by ascribing operator nature to the field
variables and imposing canonical commutation relations for the vector
field $\bf A$ and its conjugate momentum ${\bf \Pi}={\bf -D}$. These
are most easily implemented by expanding the field operators in terms
of photon annihilation and creation operators, $a_{\bf k}^{s}$ and
$a_{\bf k}^{s\: \dagger}$ respectively for a mode of  momentum $\bf k$
and polarization $s$, and demanding that the latter fulfil the
standard commutation relations
\renewcommand{\arraystretch}{1.4}
\begin{equation}
	\begin{array}{l}
	\left[ a_{\bf k}^{s}\:,\;a_{{\bf k}'}^{s'\: \dagger} \right]
	= \delta({\bf k}-{\bf k}')\:\delta_{ss'}\;, \\
	\left[ a_{\bf k}^{s}\:,\;a_{{\bf k}'}^{s'} \right] = 0\;.
	\end{array}
\label{comm}
\end{equation}\renewcommand{\arraystretch}{1}

At the same time the normal-mode expansions should be chosen such as
to diagonalize the Hamiltonian (\ref{h0}) to the Hamiltonian of a
photon field
\begin{equation}
	H_0 = \sum_s \int {\rm d}^3 {\bf k}\; \omega \: \left[
	a_{\bf k}^{s\: \dagger}\:a_{\bf k}^{s} + \frac{1}{2} \right]\;,
	\hspace*{5mm}\omega = | {\bf k} |\;.
\label{hphot}
\end{equation}

All this is achieved by the following expansion of the electric
displacement $\bf D$ and the magnetic field $\bf B$:
\renewcommand{\arraystretch}{1.5}
\begin{equation}
	\begin{array}{l}
	{\bf D}_{\rm TE} = \varepsilon\:\int{\rm d}^3 {\bf k}\;
	\displaystyle{\frac{{\rm i}\omega}{\sqrt{\omega}}}\;
	\left[ a_{\bf k}^{\rm TE}
	{\bf A}_{(1)}^{\rm TE} - {\rm H\:C\:} \right]\;, \\
	{\bf B}_{\rm TE} = \sqrt{\varepsilon}\:\int{\rm d}^3 {\bf k}\;
	\displaystyle{\frac{\omega}{\sqrt{\omega}}}\; \left[ a_{\bf k}^{\rm TE}
	{\bf A}_{(2)}^{\rm TE} + {\rm H\:C\:} \right]\;, \\
	{\bf D}_{\rm TM} = \varepsilon\:\int{\rm d}^3 {\bf k}\;
	\displaystyle{\frac{{\rm i}\omega}{\sqrt{\omega}}}\;
	\left[ a_{\bf k}^{\rm TM}
	{\bf A}_{(2)}^{\rm TM} - {\rm H\:C\:} \right]\;, \\
	{\bf B}_{\rm TM} = \sqrt{\varepsilon}\:\int{\rm d}^3 {\bf k}\;
	\displaystyle{\frac{\omega}{\sqrt{\omega}}}\; \left[ a_{\bf k}^{\rm TM}
	{\bf A}_{(1)}^{\rm TM} + {\rm H\:C\:} \right]\;. \\
	\end{array}
\label{mode}
\end{equation}\renewcommand{\arraystretch}{1}
The mode functions ${\bf A}_{(1,2)}$ are two linearly independent
solutions of the Helmholtz equation. They satisfy the Coulomb gauge
condition $\nabla\cdot{\bf A}_{(1,2)}=0$. The fields have been
decomposed into their two transverse polarizations, chosen in
spherical coordinates as the transverse electric (TE), for which the
radial component of the displacement $\bf D$ vanishes, and the
transverse magnetic (TM), for which the radial component of the
magnetic field $\bf B$ vanishes. The mode functions ${\bf A}_{(1,2)}$
and their properties are spelled out in appendix B.

As mentioned above the quantization procedure is performed at an
arbitrary but constant bubble radius $R$. This implies that although
the Hilbert space of the quantized system stays always the same, the
set of base vectors spanning it changes with $R$. A unitary
transformation from the base at radius $R$ to the one at radius $R'$
exists in principle, but is hard if not impossible to find
explicitly. So the bubble radius $R$ serves as a parameter in
traversing a continuous sequence of bases, and in a strict notation
the photon annihilation and creation operators and the photon
eigenstates should be supplemented by a label $R$. The vacuum or
ground state of the field is defined by
\[  a_{k}^s(R) \mid 0;R\rangle = 0\;,
\hspace{10mm}\langle 0;R\mid 0;R\rangle = 1\;;\]
single-photon states are written as
\[ \mid k_s;R\rangle = a_{k}^{s\:\dagger}(R) \mid 0;R\rangle\;,
\hspace{10mm}\langle k_s;R\mid k'_{s'};R\rangle = \delta(k-k')
   \:\delta_{ss'}\;;\]
two-photon states are denoted by
\begin{eqnarray} \hspace*{-20mm}&&
\mid k_s,k'_{s'};R\rangle = a_{k}^{s\:\dagger}(R)\,a_{k'}^{s'\:\dagger}(R)
\mid 0;R\rangle\;,\nonumber\\ \hspace*{-20mm}&&
\langle k_s,k'_{s'};R \mid l_p,l'_{p'};R\rangle = \delta(k-l)\,\delta_{sp}\:
\delta(k'-l')\,\delta_{s'p'}
+ \delta(k-l')\,\delta_{sp'}\:\delta(k'-l)\,\delta_{s'p}\;;\nonumber
\end{eqnarray}
and so on for all higher photon number states.

\section{TWO-PHOTON\hspace{1mm} EMISSION\hspace{1mm} IN\hspace{1mm}
FIRST-ORDER PERTURBATION THEORY}
\setcounter{equation}{0}
The evolution of the state vector $|\psi\rangle$ of the photon field
is governed by the Schr\"{o}dinger equation
\begin{equation}
	{\rm i}\,\frac{\rm d}{{\rm d}t}\,|\psi\rangle = \left[
   	H_{0}(R) + \Delta H(R,\beta) \right] |\psi\rangle
\label{schroed}
\end{equation}
where, according to the Hamiltonian (\ref{ham}) derived in appendix A,
\renewcommand{\theequation}{\arabic{section}.\arabic{equation}\alph{multieqn}}%
\setcounter{multieqn}{1}%
\begin{eqnarray}
        H_{0} &\!=&\! \frac{1}{2} \int{\rm d}^3{\bf r} \left(
        \frac{{\bf D}^2}{\varepsilon} + {\bf B}^2 \right)\;,
\label{ham0}\\
\addtocounter{equation}{-1} \addtocounter{multieqn}{1}%
        \Delta H &\!=&\! {\bf \beta} \int{\rm d}^3{\bf r}\
        \frac{\varepsilon -1}{\varepsilon}\;
        ({\bf D}\wedge{\bf B})_r \;,
\label{hint}
\end{eqnarray}\renewcommand{\theequation}{\arabic{section}.\arabic{equation}}
and $\beta\equiv\dot{R}$ is the velocity of the bubble surface.

For the present purposes antisymmetrization of the operator product in
$\Delta H$ can be dispensed with. What will be extracted from the mode
expansion of $({\bf D}\wedge{\bf B})_r$ are products of two photon
creation operators $a_k^\dagger a_{k'}^\dagger$ which induce
two-photon transitions from the vacuum. Since, however, creation
operators commute mutually, operator ordering is inessential.

To describe the sonoluminescence process by the Hamiltonian (3.2)
means to ignore variations of the refractive index due to the periodic
compression of the water in the vicinity of the bubble. This is a
crude but innocuous approximation as long as the energies of the
phonons excited in the water stay below those of the emitted photons.

Initially the photon field is in its vacuum state while the bubble is
at rest and has some radius $R$. As discussed at the end of the
preceding section, the photon eigenstates depend parametrically on the
radius $R$ of the bubble. Hence the initial condition for the state
vector $|\psi\rangle$ reads
\begin{equation}
	|\psi(t_0)\rangle = |0; R(t_0)\rangle \;.
\label{ini}
\end{equation}

The integration of the Schr\"{o}dinger equation (\ref{schroed}) poses
a non-trivial problem since standard methods of perturbation theory
cannot be applied. The Hamiltonian $\Delta H$ (\ref{hint}) cannot be
treated as an ordinary perturbation because $\Delta H$ as well as
$H_0$ depend on the parameter $R$. The established way of dealing with
slowly parameter-dependent Hamiltonians is the adiabatic approximation
\cite{pauli}. However, the standard adiabatic approximation requires
the knowledge of the complete set of eigenfunctions of the Hamiltonian
for any allowed value of the parameter. In the present case only the
eigenfunctions of part of the Hamiltonian, namely those of $H_0$, are
known. Hence what is required is a judicious combination of standard
perturbation theory and the standard adiabatic approximation; needed
is a theory that is capable of dealing both with a perturbative
interaction Hamiltonian and with a Hamiltonian depending on a slowly
varying parameter.

Following the adiabatic theory by Pauli \cite{pauli}, one starts with
the eigenvalue equation for the unperturbed Hamiltonian $H_0$, solved
for all possible values of the parameter $R$
\begin{equation}
   H_{0}(R)\, | n(R)\rangle = E_{n}(R)\, | n(R)\rangle\;,
\label{eigen}
\end{equation}
where $E_{n}(R)$ is the $n$th eigenvalue and $| n(R)\rangle$ the
corresponding eigenvector. Here $n$ is just a label; the eigenvalue
spectrum need not be discrete. In general the levels can be multiply
degenerate, so that $| n(R)\rangle$ in fact stands for a whole
subspace of orthonormal eigenvectors to the same eigenvalue
$E_{n}(R)$. Where degeneracy matters it will be explicitly indicated
by states $| n'(R)\rangle$ also belonging to $E_{n}(R)$.

Differentiating (\ref{eigen}) with respect to $R$ and calculating the
overlap with a state $\langle m(R)|$ one obtains
\[ \langle m| \frac{\partial H_{0}}{\partial R}| n\rangle
   + E_{m} \langle m| \frac{\partial}{\partial R}| n\rangle
   = \frac{\partial E_{n}}{\partial R} \langle m| n\rangle
   + E_{n} \langle m| \frac{\partial}{\partial R}| n\rangle\;.\]
Thus, provided no level-crossing occurs, ie if for $m\neq n$  is $E_{m}(R) -
E_{n}(R)$ is different from zero for all possible $R$, as it will be
the case in the present application, one has
\begin{equation}
    \langle m| \frac{\partial}{\partial R}| n\rangle =
    \frac{1}{E_{n}-E_{m}} \langle m| \frac{\partial
    H_{0}}{\partial R}| n\rangle\hspace{10mm}
    {\rm for}\ \;m\neq n\,.
\label{relat}
\end{equation}

Seeking a solution of the Schr\"{o}dinger equation (\ref{schroed}), one
expands the wave-vector $|\psi(t)\rangle$ into the eigenvectors of the
instantaneous $H_{0}(R(t))$
\begin{equation}
   |\psi(t)\rangle = \sum_{n} \hspace{-5mm} \int \hspace{2mm}
   | n(R(t))\rangle\langle n(R(t))|\psi(t)\rangle\;.
\label{expan}
\end{equation}
Then the Schr\"{o}dinger equation (\ref{schroed}) becomes
\[ \sum_{n} \hspace{-5mm} \int \hspace{2mm} \left[ {\rm i} \left(
   \frac{\partial}{\partial R}| n\rangle\right)
   \frac{\partial R}{\partial t} \langle n|\psi\rangle
   + {\rm i} | n\rangle \left( \frac{\partial}{\partial t}
   \langle n|\psi\rangle \right) \right]
   = \sum_{n} \hspace{-5mm} \int \hspace{2mm} \left[ E_{n}
   | n\rangle\langle n|\psi\rangle + \Delta H | n\rangle\langle
   n|\psi\rangle \right]\;, \]
which by taking the scalar product with an eigenstate $\langle m|$ is
turned into
\[ \sum_{n} \hspace{-5mm} \int \hspace{2mm} {\rm i}\beta
   \langle n|\psi\rangle \langle m|\frac{\partial}{\partial R}|
   n\rangle + {\rm i} \frac{\partial}{\partial t} \langle m|\psi\rangle
   = E_{m} \langle m|\psi\rangle + \sum_{n} \hspace{-5mm} \int
   \hspace{2mm} \langle m| \Delta H | n\rangle\langle
   n|\psi\rangle \;.\]
{}From here, application of the relation (\ref{relat}) yields
\begin{eqnarray} &&\hspace*{-20mm}
   {\rm i} \frac{\partial}{\partial t} \langle m|\psi\rangle - E_{m}
   \langle m|\psi\rangle + {\rm i}\beta \sum_{m'(E_{m'}=E_{m})}
   \hspace{-12mm} \int
   \hspace{9mm} \langle m'|\psi\rangle \langle m|
   \frac{\partial}{\partial R}| m'\rangle
   \nonumber\\
   &&= -{\rm i}\beta \sum_{n\;(n\neq m)} \hspace{-8.5mm} \int^{\ \prime}
   \hspace{5.5mm} \frac{\langle n|\psi\rangle}{E_{n}-E_{m}}
   \langle m| \frac{\partial H_{0}}{\partial R}| n\rangle +
   \sum_{n} \hspace{-5mm} \int \hspace{2mm} \langle m| \Delta H |
   n\rangle \langle n|\psi\rangle\;,\nonumber
\end{eqnarray}
where the sum over $m'$ takes states degenerate with $m$ into account. Since
eventually the transition probability $|\langle m|\psi\rangle|^{2}$ and
not the transition amplitude $\langle m|\psi\rangle$ will be of physical
interest, one can gauge away the second term on the left-hand side of the
above equation by defining
\begin{equation}
   \langle m|\psi\rangle = c_{m}\,\exp\left[ -{\rm i}\int_{t_{0}}^{t}
   {\rm d}\tau\; E_{m}(\tau)\right]\;;
\label{defc}
\end{equation}
one finds
\begin{eqnarray} &&\hspace*{-20mm}
   \frac{\partial c_{m}}{\partial t} + \beta \sum_{m'(E_{m'}=E_{m})}
   \hspace{-12mm} \int
   \hspace{9mm} c_{m'} \langle m|\frac{\partial}{\partial R}| m'\rangle
   \nonumber\\
   &&\hspace*{-10mm}= \beta\sum_{n\;(n\neq m)} \hspace{-8.5mm} \int^{\ \prime}
   \hspace{5.5mm} \frac{c_{n}}{E_{n}-E_{m}} \langle m|
   \frac{\partial H_{0}}{\partial R}| n\rangle \exp\left[ {\rm i}
   \int_{t_{0}}^{t} {\rm d}\tau\; (E_{m}-E_{n})\right]\nonumber\\
   &&\hspace{15mm}-{\rm i} \sum_{n} \hspace{-5mm} \int \hspace{2mm}
   c_{n} \langle m|
   \Delta H| n\rangle \exp\left[ {\rm i} \int_{t_{0}}^{t} {\rm d}\tau\;
   (E_{m}-E_{n})\right]\;,
\label{key}
\end{eqnarray}
which is the key formula for the present approximation method. If it
were not for the term containing $\Delta H$, this expression would
lead to merely the standard adiabatic approximation (cf. for instance
ref. \cite{schiff}). It should be noted that both terms on the
right-hand side of (\ref{key}) are of the same order, namely
$\beta^{1}$; the matrix element of $\partial H_{0}/\partial R$ is
multiplied by $\beta$, and $\Delta H$ is itself of order $\beta$.

The initial condition (\ref{ini}) for the wave function $|\psi\rangle$
translates into the following initial conditions for the coefficients
$c_m(t)$ defined by (\ref{defc})
\begin{equation}
   c_{0}(t_{0}) = 1\;, \hspace{10mm}c_{m\neq 0}(t_{0}) = 0\;.
\label{cini}
\end{equation}
As soon as the bubble starts moving, ie the interface velocity
$\beta(t>t_{0})$ becomes different from zero, the rate of change of
the $c_{m}$ is non-zero, as described by eq (\ref{key}). For times
$t>t_{0}$ one has $c_{0}(t) \approx 1$ and $c_{m\neq 0}(t) =
O\,(\beta)$ or higher. Hence, working only to first order in $\beta$,
one has to retain only the vacuum state in the summation over $n$ on
the right-hand side of eq (\ref{key}).

Handling the time-dependence of the energy eigenvalues $E_n$ requires
special care. For a cavity, like the bubble in the present problem,
one has two limiting cases. The first is that the cavity walls are
very poor reflectors; then it is convenient to label the cavity modes
by wavenumber or energy as these are adiabatically conserved. However,
if the cavity has a very high Q value, i.e. is close to perfectly
reflecting, the number of nodes of the eigenfunction will be the
adiabatically conserved quantity and not the wavenumber. The
probability of reflection from an interface of a medium of refractive
index $n$ with the vacuum is given by $(n-1)^2/(n+1)^2$. For an
air-water interface in the visible spectrum where $n=1.3$ this leads
to a reflection probability of less than 2\%. Hence the bubble is a
poor-quality cavity and the energy eigenvalues are adiabatically
conserved. This justifies the use of wavenumbers for the labelling of
eigenstates.

Since $H_{0}$, and hence $\partial H_{0}/\partial R$, and $\Delta H$
are quadratic in the fields, the only transitions they can induce from
the initial vacuum state lead to two-photon states, whence to first
order in $\beta$ \cite{fn2} the system of differential equations
(\ref{key}) for the coefficients $c_{m}$ reduces to
\renewcommand{\theequation}{\arabic{section}.\arabic{equation}\alph{multieqn}}%
\setcounter{multieqn}{1}%
\begin{eqnarray} &{\displaystyle
   \frac{\partial c_{0}}{\partial t} + \beta \langle 0; R|
   \frac{\partial}{\partial R}|0; R\rangle = - {\rm i}\langle 0; R|
   \Delta H |0; R\rangle\;,} &
\label{c0}\\
\addtocounter{equation}{-1} \addtocounter{multieqn}{1}%
   &\hspace*{-15mm}{\displaystyle
   \frac{\partial c_{kk'}}{\partial t} = \beta \,\frac{1}{\omega +\omega'}\,
   \langle k,k'; R| \frac{\partial H_{0}}{\partial R}|0; R\rangle\,
   {\rm e}^{{\rm i}(\omega +\omega')(t-t_{0})}
   -{\rm i} \langle k,k'; R| \Delta H |0; R\rangle\,
   {\rm e}^{{\rm i}(\omega +\omega')(t-t_{0})}\;.} &
\label{ckk}
\end{eqnarray}\renewcommand{\theequation}{\arabic{section}.\arabic{equation}}
The first equation is of no special interest,
but the second equation will readily provide the solution of the problem
posed --- the perturbative description of the photon creation by the moving
bubble surface. According to (\ref{defc}) the transition amplitude from the
vacuum into a two-photon state is given by
\begin{equation}
   \langle k,k'; R|\psi\rangle = c_{kk'}(t)\;{\rm e}^{-{\rm i}(\omega
   +\omega')(t-t_{0})}\;,
\label{triv}
\end{equation}
and the initial vacuum evolves into the state
\begin{equation}
   |\psi\rangle = |0; R\rangle + \frac{1}{2} \int_{-\infty}^{\infty} {\rm d}k
   \int_{-\infty}^{\infty} {\rm d}k'\; c_{kk'}(t)\;{\rm e}^{-{\rm i}(\omega
   +\omega')(t-t_{0})}\, |k,k'; R\rangle\;.
\label{decomp}
\end{equation}
The factor 1/2 takes care of the identical photon states $|k,k'; R\rangle
=|k',k; R\rangle$, not to be double counted.

It remains to evaluate the two matrix elements in eq (\ref{ckk}). With
$H_{0}$ as in (\ref{ham0}) one finds for the first
\begin{equation}
	\langle k,k'; R|\, \frac{\partial H_0}{\partial R} |0; R\rangle
	= \frac{1}{2} \left( 1-\frac{1}{n^2} \right) R^2 \oint {\rm d}
	\Omega\:  \langle k,k'; R|\;D_r^2 + n^2 {\bf E}_\parallel^2\;
	|0; R\rangle\;,
\label{matr1}
\end{equation}
in the obvious notation ${\bf E}_\parallel=(E_\theta,E_\phi)$.
Deriving this, one should bear in mind that $H_0$ depends on $R$ both
through $\varepsilon(R)$ and through the discontinuity of ${\bf
D}_\parallel=(D_\theta,D_\phi)$.

With $\Delta H$ as in (\ref{hint}) the second matrix element in
(\ref{ckk}) reads
\begin{equation}
	\langle k,k'; R|\,\Delta H |0; R\rangle =
	\beta\; \left(1-\frac{1}{n^2}\right) \int_{r\geq R} {\rm d}^3
	{\bf r}\; \langle k,k'; R|\,
	({\bf D} \wedge {\bf B})_r |0; R\rangle\;.
\label{mhint}
\end{equation}
As photon states are eigenstates of $H_0$, one can write
\begin{eqnarray}
	{\rm i}\,(\omega + \omega')\, \langle k,k'; R|\,
	({\bf D} \wedge {\bf B})_r |0; R\rangle &\! =&\!
	\langle k,k'; R|\,{\rm i}\:[ H_0 ,
	({\bf D} \wedge {\bf B})_r ] |0; R\rangle
	\nonumber\\ &\! =&\!
	\langle k,k'; R|\,\frac{\partial}{\partial t}\:
	({\bf D} \wedge {\bf B})_r |0; R\rangle\;.
\label{help}
\end{eqnarray}
to first order in $\beta$.
In further manipulating this expression, one can make use of the
classical energy-momentum conservation law in a bulk dielectric
\begin{equation}
	\frac{\partial}{\partial t}\:({\bf D} \wedge {\bf B})_i
	+ \nabla_j T^{ij} = 0\;,
\label{cons}
\end{equation}
where the stress-tensor in the medium is given by
\begin{equation}
	T^{ij} = - \frac{D_i D_j}{\varepsilon} - B_i B_j + \frac{1}{2}
	\delta_{ij} \left( \frac{{\bf D}^2}{\varepsilon} + {\bf B}^2 \right)\;.
\label{Tij}
\end{equation}
Since the photon field is a non-self-interacting field in Minkowski
space, there is no doubt that this conservation law is valid also
quantally. Using this and the above relation (\ref{help}), one
can rewrite (\ref{mhint})
\[	\langle k,k'; R|\,\Delta H |0; R\rangle = {\rm i}
	\frac{\beta}{\omega+\omega'}
	\left(1-\frac{1}{n^2}\right) \int_{r\geq R} {\rm d}^3 {\bf r}\;
	\langle k,k'; R|\,\nabla_j T^{rj} |0; R\rangle\;.
\]
Applying Gauss' theorem leads to
\begin{eqnarray}
	\langle k,k'; R|\,\Delta H |0; R\rangle &&\hspace*{-6mm}
	= - {\rm i} \frac{\beta R^2}{\omega+\omega'}
	\left(1-\frac{1}{n^2}\right) \oint {\rm d} \Omega\;
	\langle k,k'; R|\,T^{rr} |0; R\rangle
\nonumber\\ &&\hspace*{-30mm}=
	- {\rm i} \frac{\beta R^2}{\omega+\omega'}
	\left(1-\frac{1}{n^2}\right) \oint {\rm d} \Omega\;
	\langle k,k'; R|\;\frac{1}{2}\left(n^2 {\bf E}_\parallel^2
	- \frac{D_r^2}{n^2} + {\bf B}_\parallel^2 - B_r^2 \right)
	|0; R\rangle\;.
\label{matr2}
\end{eqnarray}

Inserting (\ref{matr1}) and (\ref{matr2}) into the expression
(\ref{ckk}) one finds for the time-derivative of the transition
amplitude between the vacuum and a two-photon state $|k,k'\rangle$
\[	\frac{\partial c_{kk'}}{\partial t} =
	\left(1-\frac{1}{n^2}\right) \frac{\beta\,R^2}{2}\,
	\frac{{\rm e}^{{\rm i}(\omega+\omega')(t-t_0)}}{\omega +\omega'}
	\oint {\rm d} \Omega\;
	\langle k,k'; R| \left(1+\frac{1}{n^2}\right)
	D_r^2 - {\bf B}_\parallel^2 + B_r^2\, |0; R\rangle\;.
\]
Comparison with the force operator ${\cal F}_r$ given by eq. (\ref{force})
uncovers the relation of the photon creation to the pressure on the
bubble; $\partial c_{kk'}/\partial t$ can be re-expressed as
\begin{equation}
	\frac{\partial c_{kk'}}{\partial t} = -
	\frac{\beta}{\omega +\omega'}\:
	{\rm e}^{{\rm i}(\omega+\omega')(t-t_0)}\:
	\langle k,k'; R|\;{\cal F}_r\,|0; R\rangle\;.
\label{dckk}
\end{equation}
This is a truly remarkable result as it exposes the fluctuations of
the radiation pressure as the origin of the photon-pair creation. As
described by eq. (\ref{statforce}) for the mean-square deviation of
the force, the fluctuations on a stationary mirror are tied in with
the excitation of virtual two-photon states. The (non-uniform) motion
of a mirror or a dielectric interface makes these virtual states
become real, which has been explained in the two paragraphs following
eq. (\ref{statforce}) and is shown manifestly by eq. (\ref{dckk}). The
fluctuation-dissipation theorem underlies this connection; it however
predicts only the dissipative force acting on the moving interface and
not the photon-creation amplitude. Hence, the fluctuation-dissipation
theorem cannot supersede the above derivation, but is nevertheless a
useful check on it \cite{CEfluc}.

The integration of eq. (\ref{dckk}) is complicated by the fact that
the force matrix-element depends parametrically on $R$ and therefore
on time. Formally, the transition amplitude reads
\begin{equation}
	c_{kk'}(t) = - \frac{1}{\omega +\omega'}\: \int_{t_0}^{t}
	{\rm d}\tau\;\beta(\tau)\:{\rm e}^{{\rm i}(\omega+\omega')(\tau-t_0)}\:
	\langle k,k'; R(\tau)|\;{\cal F}_r\,|0; R(\tau)\rangle\;.
\label{intc}
\end{equation}
The matrix elements of the force operator, which is given by
(\ref{force}), are calculated by expanding the $\bf D$ and $\bf B$
fields into normal modes as detailed in section 2 and appendix B. An
unsophisticated calculation yields
\renewcommand{\theequation}{\arabic{section}.\arabic{equation}\alph{multieqn}}%
\setcounter{multieqn}{1}%
\begin{eqnarray}
	\langle k_{\rm TE}, k_{\rm TE}';R|\; {\cal F}_{r}\,|0; R(\tau)\rangle
	&\!=&\! \frac{n}{2\pi} \left( 1 - \frac{1}{n^2} \right)
	\sqrt{\omega\omega'}\; \sum_{\ell,m}\;
	{\cal S}^{{\rm TE}\;-1}_{\ell}(k)\;
	{\cal S}^{{\rm TE}\;-1}_{\ell}(k')\; \frac{(-1)^\ell}{kk'}
\nonumber\\ && \hspace*{-10mm} \times
	\left\{
	\ell(\ell+1)\: j_{\ell}(kR)\:j_{\ell}(k'R) + \left[ kR\;j_{\ell}(kR)
	\right]^{\prime} \left[ k'R \;j_{\ell}(k'R) \right]^{\prime}
	\right\}
\nonumber\\ && \hspace*{-10mm} \times\
	Y_{\ell}^{m}(\hat{\bf k})\:Y_{\ell}^{m\ast}(\hat{\bf k}')\;,
\label{FTE}\\
\addtocounter{equation}{-1} \addtocounter{multieqn}{1}%
	\langle k_{\rm TM}, k_{\rm TM}';R|\; {\cal F}_{r}\,|0; R(\tau)\rangle
	&\!=&\! \frac{n}{2\pi} \left( 1 - \frac{1}{n^2} \right)
	\sqrt{\omega\omega'}\; \sum_{\ell,m}\;
	{\cal S}^{{\rm TM}\;-1}_{\ell}(k)\;
	{\cal S}^{{\rm TM}\;-1}_{\ell}(k')\; \frac{(-1)^\ell}{kk'}
\nonumber\\ && \hspace*{-10mm} \times
	\left\{
	kk'R^2 - \ell(\ell+1) \right\}\: j_{\ell}(kR)\:j_{\ell}(k'R)\;
	Y_{\ell}^{m}(\hat{\bf k})\:Y_{\ell}^{m\ast}(\hat{\bf k}')\;,
\label{FTM}\\
\addtocounter{equation}{-1} \addtocounter{multieqn}{1}%
	\langle k_{\rm TM}, k_{\rm TM}';R|\; {\cal F}_{r}\,|0; R(\tau)\rangle
	&\!=&\! 0\;,
\end{eqnarray}\renewcommand{\theequation}{\arabic{section}.\arabic{equation}}
where the normalization constants ${\cal S}^{{\rm TE,TM}\;-1}_{\ell}(k)$
are as given in eqs. (\ref{STEM}) and (\ref{NandD}).

The probability of creating a photon pair in the mode $|k,k'\rangle$
from the initial vacuum state is given by the modulus square of the
amplitude $c_{kk'}$, eq. (\ref{intc}), featuring the above matrix
elements of the force operator. The state of the photon field is
specified by eq. (\ref{decomp}). Hence $c_{kk'}$ carries all
information one needs to determine the expectation values of all
interesting observables, especially the total radiated energy and the
spectral density, which are the subject of the following section.

\section{RADIATED ENERGY AND SPECTRAL DENSITY}
\setcounter{equation}{0}
As $|c_{kk'}(t)|^{2}$ is the probability of creating a photon pair in
the mode $|k,k'\rangle$, the total energy of the photons radiated by
the bubble interface during one acoustic cycle reads
\begin{equation}
	{\cal W} = \frac{1}{2}\,\int_0^T {\rm d}t\,
	\int_{-\infty}^{\infty}{\rm d}^3 {\bf k}\,
   	\int_{-\infty}^{\infty}{\rm d}^3 {\bf k}'\;(\omega + \omega')\;
   	|\,c_{kk'}(t)|^{2}\;,
\end{equation}
where $T$ is the period of the sound field. Inserting $c_{kk'}(t)$
from (\ref{intc}) and (3.21) one obtains for $\cal W$
\begin{eqnarray} &&\hspace*{-12mm}
	{\cal W} = \frac{(n^2-1)^2}{8\pi^2n^2} \int_0^\infty {\rm d}
	\omega \int_0^\infty {\rm d} \omega'\;
	\frac{\omega\omega'}{\omega+\omega'}\;
        \int_0^T {\rm d}\tau \int_0^T {\rm d}\tau'\,
	\beta(\tau)\beta(\tau')\;
	{\rm e}^{{\rm i}(\omega+\omega')(\tau-\tau')}
\nonumber\\ && \hspace*{90mm} \times\:
	\Im(k,k',R(\tau),R(\tau'))\;,
\label{fullW}
\end{eqnarray}
with the auxiliary function $\Im$ defined as
\begin{eqnarray} &&\hspace*{-12mm}
	\Im(k,k',R(\tau),R(\tau'))=
	\sum_{\ell=1}^{\infty}\:(2\ell+1)
\nonumber\\ && \hspace*{-4mm}\times
	\left\{ {\cal S}^{{\rm TE}\;-1}_{\ell}(kR(\tau))\,
	{\cal S}^{{\rm TE}\;-1}_{\ell}(k'R(\tau))\,
	{\cal S}^{{\rm TE}\,\ast\;-1}_{\ell}(kR(\tau'))\,
	{\cal S}^{{\rm TE}\,\ast\;-1}_{\ell}(k'R(\tau'))\right.
\nonumber\\ && \times
	\left[ \ell(\ell+1)\,j_{\ell}(kR(\tau))\,j_{\ell}(k'R(\tau))
	+ \left[ kR(\tau) \;j_{\ell}(kR(\tau)) \right]^{\prime}
	\left[ k'R(\tau) \;j_{\ell}(k'R(\tau)) \right]^{\prime}\right]
\nonumber\\ && \times
	\left[ \ell(\ell+1)\,j_{\ell}(kR(\tau'))\,j_{\ell}(k'R(\tau'))
	+ \left[ kR(\tau') \;j_{\ell}(kR(\tau')) \right]^{\prime}
	\left[ k'R(\tau') \;j_{\ell}(k'R(\tau')) \right]^{\prime}\right]
\nonumber\\ && \hspace*{-4mm}+
	\;{\cal S}^{{\rm TM}\;-1}_{\ell}(kR(\tau))\,
	{\cal S}^{{\rm TM}\;-1}_{\ell}(k'R(\tau))\,
	{\cal S}^{{\rm TM}\,\ast\;-1}_{\ell}(kR(\tau'))\,
	{\cal S}^{{\rm TM}\,\ast\;-1}_{\ell}(k'R(\tau'))
\nonumber\\ && \times
	\left[ kk'R^2(\tau) - \ell(\ell+1) \right]
	\left[ kk'R^2(\tau') - \ell(\ell+1) \right]
\nonumber\\ && \times
	\left. j_{\ell}(kR(\tau))\,j_{\ell}(k'R(\tau))\,
	j_{\ell}(kR(\tau'))\,j_{\ell}(k'R(\tau'))\right\}\;,
\label{helpR}
\end{eqnarray}
and the normalization factors ${\cal S}^{{\rm TE,TM}\;-1}_{\ell}(k)$
as given by eqs. (\ref{STEM}) and (\ref{NandD}).

A spectrometer measures the single-photon spectrum, which by symmetry
is isotropic for a spherical bubble. The quantity of interest is
therefore the angle-integrated spectral density radiated during one acoustic
cycle,
\begin{equation}
	{\cal P}(\omega) = \omega^3\int_0^T {\rm d}t
	\oint {\rm d} \Omega_{\bf k}\;
	\int_{-\infty}^{\infty}{\rm d}^3 {\bf k}'\;
   	|\,c_{kk'}(t)|^{2}\;,
\label{spectr}
\end{equation}
which becomes
\begin{eqnarray} &&\hspace*{-12mm}
	{\cal P}(\omega) = \frac{(n^2-1)^2}{4\pi^2n^2}\;
	\omega^2 \int_0^\infty {\rm d} \omega'\;
	\frac{\omega'}{(\omega+\omega')^2}\;
        \int_0^T {\rm d}\tau \int_0^T {\rm d}\tau'\,
	\beta(\tau)\beta(\tau')\;
	{\rm e}^{{\rm i}(\omega+\omega')(\tau-\tau')}
\nonumber\\ && \hspace*{90mm} \times\:
	\Im(k,k',R(\tau),R(\tau'))\;.
\label{fullP}
\end{eqnarray}

The main difficulty in calculating the radiated energy $\cal W$ and
the spectrum ${\cal P}(\omega)$ is the evaluation of the auxiliary
function $\Im$. To find an analytical approximation for $\Im$ is a
rather laborious task.

First, note that both $\cal W$ and ${\cal P}(\omega)$,
eqs. (\ref{fullW}) and (\ref{fullP}), contain a prefactor $(n^2-1)^2$
multiplying $\Im$, so that they vanish in the limit $n\rightarrow 1$,
as they should; in this limit there is no dielectric interface to
produce radiation. This justifies an expansion around $n=1$ in $\Im$,
which to first order does nothing but reduce the normalization factors
${\cal S}^{{\rm TE,TM}\;-1}_{\ell}$ in (\ref{helpR}) to 1, by virtue
of eq. (\ref{Born}); otherwise $\Im$ is independent of $n$.

In the present application the arguments of the Bessel functions in
$\Im$ are generally greater than 1, partly appreciably much greater,
so that expansions for Bessel functions of small arguments are of no
use here. As the summation over the index $\ell$ runs up to infinity,
the sum will be dominated by terms for which argument and index of
each of the Bessel functions are comparable in magnitude. Hence
Debye's uniform asymptotic expansion has to be employed (cf. \cite{AS}
formulae 9.3.3 and 9.3.7). So, for instance, one obtains in the regime
$x\geq(\ell+1/2)$
\[ 	j_{\ell}(x)\longrightarrow
	\frac{1}{\nu}\:\frac{1}{\sqrt{\sec\beta\tan\beta}} \: \cos(\nu
	\tan\beta -\nu\beta -\frac{\pi}{4})
\]
and
\[ 	[x\;j_{\ell}(x)]^\prime\longrightarrow
	-\sqrt{\frac{\tan\beta}{\sec\beta}} \: \sin(\nu
	\tan\beta -\nu\beta -\frac{\pi}{4})\;,
\]
where the abbreviations $\nu\equiv(\ell+1/2)$ and $x\equiv\nu\sec\beta$
have been introduced.

With the help of the above asymptotic approximations and by turning
the summation over $\ell$ into an integration one can derive that
$\Im$ behaves approximately like $kk'R(\tau)R(\tau')$ in the
short-wavelengths regime, i.e. when the photon wavelengths are shorter
than the minimum bubble radius. Numerical investigation of the
behaviour of $\Im$ confirms this and yields
\begin{equation}
	\Im \sim 1.16\; kk'R(\tau)R(\tau')\;.
\label{approx}
\end{equation}
Employing this approximation one can integrate the expression
(\ref{fullW}) for $\cal W$ and obtains after a short calculation:
\begin{equation}
	{\cal W} = 1.16\:\frac{(n^2-1)^2}{n^2}\,\frac{1}{480\pi}
	\int_{0}^{T} {\rm d}\tau\; \frac{\partial^5 R^2(\tau)}{\partial
	\tau^5}\,R(\tau)\beta(\tau)\;.
\label{Wres}
\end{equation}
One of the interesting consequences of this result is that the
dissipative force acting on the moving dielectric interface can be
seen to behave like $R^2\beta^{(4)}(t)$ (+ terms with lower
derivatives of $\beta$). This dependence tallies with results of
calculations for frictional forces on moving perfect mirrors (see esp.
\cite{paulo}); the dissipative part of the radiation pressure on a
moving dielectric or mirror is proportional to the fourth derivative of
the velocity.

The expression (\ref{Wres}) indicates also that any discontinuity in
$\beta^{(3)}(t)$ or lower derivatives of $\beta$ unavoidably leads to
a divergence in $\cal W$ (and also in the spectral density ${\cal
P}(\omega)$). Especially, one is not permitted to assume a
step-function profile for $R(t)$ during the collapse of the bubble,
since this would give the physically meaningless result of infinite
photon production, which is not salvageable by a cut-off or any other
artificial regularization
\cite{jschw}.

What produces the massive burst of photons from a collapsing
sonoluminescent bubble is the turn-around of the velocity at the
minimum radius of the bubble. There the velocity rapidly changes sign,
from collapse to re-expansion of the bubble. This means that the
acceleration is peaked at this moment and so are higher derivatives of
the velocity.

In order to estimate the total energy radiated during one acoustic
cycle, one can use the approximation (\ref{approx}) to rewrite the
expression (\ref{fullW}) as
\begin{equation}
	{\cal W} = 1.16\: \frac{(n^2-1)^2}{960\,\pi^2n^2} \,
	\int_0^\infty {\rm d}\Omega\; \Omega^4\: \left|
	\int_0^T {\rm d}\tau \; \frac{\partial R^2(\tau)}{\partial\tau}
	\;{\rm e}^{{\rm i}\Omega\tau} \right|^2
\label{simpleW}
\end{equation}
where $\Omega$ is the sum of the photon frequencies in a pair. As
$\cal W$ is a functional of the time-dependent radius $R(t)$, one has
to model $R(t)$ appropriately in order to be able to obtain a number
for $\cal W$. At the collapse of the bubble the function $R(t)$ has a
sharp dip; hence it is reasonable to adopt the model profile
\begin{equation}
	R^2(t) = R_0^2 - \left(R_0^2 - R_{\rm min}^2\right) \,
	\frac{1}{(t/\gamma)^2 + 1}
\label{modR}
\end{equation}
for the time-dependence of the bubble radius. Figure 1 illustrates
$R^2(t)$ for various values of the parameter $\gamma$ which
describes the timescale of the collapse and re-expansion process.
The shorter $\gamma$ the faster is the turn-around of the velocity at
minimum radius and the more violent is the collapse. In the figure
$\gamma$ is half of the width of the dip halfway between $R_0^2$ and
$R_{\rm min}^2$. In this simple model the total radiated energy
(\ref{simpleW}) reads in SI units
\begin{equation}
	{\cal W} = 1.16\: \frac{3\,(n^2-1)^2}{512\,n^2} \,
	\frac{\hbar}{c^4\gamma^{5}}\,
	\left( R_0^2 - R_{\rm min}^2\right)^2\;.
\label{exW}
\end{equation}
Experimental data on sonoluminescent bubbles \cite{miemeas} suggest
that $R_0 \sim 10\mu$m and $R_{\rm min} \sim 0.5\mu$m are sensible
values to assume. With $n\sim 1.3$ one obtains
\begin{equation}
	{\cal W} = 1.8\cdot 10^{-13}\,{\rm J}\ \ {\rm for}\
	\gamma\sim 1{\rm fs}\,,
\label{numW}
\end{equation}
which corresponds roughly to the experimentally observed amount of
energy per burst. Calculating the radiated spectral density in the
same model gives
\begin{equation}
	{\cal P}(\omega) = 1.16\: \frac{(n^2-1)^2}{64\,n^2} \,
	\frac{\hbar}{c^4\gamma}\,
	\left( R_0^2 - R_{\rm min}^2\right)^2\:
	\omega^3\:{\rm e}^{-2\gamma\omega}\;.
\label{specdens}
\end{equation}
This is one of the most important end-results of this calculation, as
it exhibits the same $\omega$ dependence as black-body radiation.
Equating the exponent in (\ref{specdens}) to $\hbar\omega/kT$ one
derives that a turn-around time $\gamma$ of 1fs corresponds to a
temperature of around 4000K. This is however just a very crude
estimate, as a lot of simplifications and approximations have been
made in proceeding from (\ref{fullW}) to (\ref{specdens}); in general,
the functional dependence of $\cal P$ on $\omega$ will not be as
simple, although its overall behaviour is as characterized by
eq. (\ref{specdens}).

The major problem in the above derivation is that the approximation
(\ref{approx}) is good only at photon wavelengths that are smaller
than the bubble radius $R$, i.e. when all products $kR$ and $k'R$ are
greater than 1. Once $kR$ reaches down to the order of 1 or even
below, one has to expect resonances of the photon wavelengths with the
bubble radius. An exploration of any such resonance effects requires
taking into account the full $kR$ dependence of the auxiliary integral
$\Im$ in eq. (\ref{helpR}), which is fairly difficult even numerically.

The results of a computer simulation of a model like the one specified
by (\ref{modR}) with $R_0=45\mu$m, $R_{\rm min}=3\mu$m, and
$\gamma=10$fs are shown in figures 2 and 3. Comparing these with the
values of $\cal P$ predicted by eq. (\ref{specdens}), one sees that
the numerical results show an enhancement of about a factor 1000 in
$\cal P$ relative to the analytical approximation. This is due to
resonant behaviour in (\ref{modR}), since $R_{\rm min}$ is not any
more appreciably much larger than the wavelength of the observed
light. Numerical studies in the regime
\renewcommand{\arraystretch}{0.5} $kR_{\rm min} \begin{array}{c} < \\
\sim \end{array} 1$ \renewcommand{\arraystretch}{1.0}
are hindered by substantial expense in
computation time; work by the present author is in progress
\cite{progress}. One can expect to see an even greater enhancement in
$\cal P$ over the predictions of eq. (\ref{specdens}) for what are
believed to be realistic values of $R_{\rm min}$, i.e. values around
$0.5\mu$m. Thus, one can presumably substantially relax the
requirement on how small the turn-around time $\gamma$ between
collapse and re-expansion of the bubble has to be in order for the
present theory to yield the experimentally observed number of photons.
The crude model that led to the estimate (\ref{numW}) would demand
$\gamma$ to be as short as 1fs, if it were to account for the
experimental data for the sonoluminescence of an air bubble in
water. The numerical calculation resulting in the figures 2 and 3
suggests that $\gamma\sim 10$fs is fully sufficient; calculations for
$R_{\rm min}$ smaller than $3\mu$m will most likely require merely
$\gamma\sim 100{\rm fs}\ldots10{\rm ps}$, which is closer to what one
expects this timescale to be on physical grounds.

The features seen in the spectrum in figure 2 seem to be due to
resonances between photon wavelengths and bubble size, but again,
extensive numerical studies are needed to explore them
\cite{progress}. Comparison of the figures 2 and 3 makes it very clear
that plotting data over photon wavelength rather than frequency tends
to conceal such features; it might therefore be beneficial to plot
experimental data over photon frequency as well.

Given the time-dependence of the bubble radius $R(t)$, one can in
principle, technical difficulties aside, predict the sonoluminescence
spectrum radiated by the bubble from eqs. (\ref{fullP}) and
(\ref{helpR}). The next section summarizes the questions answered by
the theory of quantum vacuum radiation and spells out some of the as
yet unanswered ones.

\section{SUMMARY AND CRITICAL REFLECTION}
 \setcounter{equation}{0}
\subsection{Successfully resolved issues}
\subsubsection{Results\hspace{1mm} for\hspace{1mm} the\hspace{1mm}
spectral\hspace{1mm} density\hspace{1mm} and\hspace{1mm} restrictions
\hspace{1mm} on\hspace{1mm} the\hspace{1mm} turn-around time $\gamma$}
The results of the previous section seem in concord with the
experimentally observed facts in sonoluminescence. Knowing the
time-dependence of the bubble radius $R(t)$ one can evaluate the
radiated spectral density from eqs. (\ref{fullP}) and
(\ref{helpR}). The analytical estimate (\ref{numW}) for the total
energy radiated during one acoustic cycle and the results of numerical
calculations presented in figs. 2 and 3 show good qualitative and
quantitative agreement with the experimental data \cite{spec}. The
function $R(t)$ has been modelled by a dip, as written down in
eq. (\ref{modR}) and made visual in fig. 1. The width of this dip is
characterized by the important parameter $\gamma$, whose physical
significance is that it is a measure of the time-scale of the
turn-around of the velocity between the collapse and the re-expansion
of the bubble. To model the experimentally observed data on the basis
of the numerical calculations worked through so far, this turn-around
has to happen in about 10fs. Further numerical analysis of
eqs. (\ref{fullP}) and (\ref{helpR}) in the regime where the radiated
photon wavelengths get in resonance with the bubble size can be
anticipated to relax this requirement to a turn-around time of roughly
100fs$\ldots$10ps, which seems quite realistic. Unfortunately, it is
notoriously difficult to determine $\gamma$ experimentally, as it is
arduous to measure the bubble radius close to the collapse; the laser
light that is used to determine the bubble radius by means of fitting
the scattering data to the Mie theory, is then most likely reflected
by the shock-wave front propagating through the water rather then by
the actual bubble surface \cite{miemeas}. In addition, one might also
have to take into account that the bubble shape deviates, perhaps even
substantially, from spherical. Although this does not affect the
present theory of the radiation mechanism to any discernible extent,
since the spectrum of the vacuum fluctuations is known to be affected
by the shape only to higher order \cite{CEdf2}, it will noticeably
alter the light-scattering properties so that the Mie scattering
theory is no longer applicable. Nevertheless, the currently available
experimental data seem to suggest that $\gamma$ lies somewhere in the
interval 100fs$\ldots$10ps\cite{miemeas}.

\subsubsection{Thermal properties of the spectral density}
The similarity of the observed photon spectrum to a black-body
spectrum has its origin in the fact that the photon radiation emerges
in coherent pairs. To obtain the single-photon spectrum, which is the
one measured in spectral analyses, one has to trace over one photon in
the pair, as done by integrating over $k'$ in eq. (\ref{spectr}). This
process of tracing is known to engender thermal properties of the
single-photon spectrum, even though the original two-photon state was
a pure state and one has dealt with zero-temperature quantum field
theory throughout \cite{knight}. In other words, it is the particular
correlations within the radiated photon pairs that engender the
thermal-like properties of the spectrum, and thermal processes are
completely independent of this and in the case of sonoluminescence
presumably of no significance whatsoever.

\subsubsection{Features in the spectrum}
Features in the otherwise smooth experimental spectra can be explained
by a combination of three things:

{\em (i)} Resonance effects in the radiation mechanism occur when the
size of the bubble $R(t)$ and the photon wavelengths $\lambda$ are of
the same order of magnitude. The features seen in the numerically
calculated spectral density in figure 2 are due to higher-order
resonances of this kind where the $R$ is an integer multiple of
$\lambda$; for direct resonances one expects much stronger effects.
General predictions for such features can hardly be made as their
detailed qualities are determined by the time-dependence of $R$, which
leads over to the next point.

{\em (ii)} Choosing gases other than air for the bubble contents
leads to a modified dynamics of the bubble surface, as gas
solubilities in water vary; but even a slightly different $R(t)$
around the collapse will change the structure of the resonance effects
between $\lambda$ and $R$, which is why one expects a strong
dependence of the sonoluminescence spectra on the gas that saturates
the water, as indeed observed in experiments
\cite{fareast,noble}.

{\em (iii)} The experimentally seen features in the spectrum might just
as well be caused by dispersion, i.e. the dependence of the refractive
index $n$ on the photon frequency $\omega$. As long as $n$ depends
only weakly on $\omega$, one can replace $n$ by $n(\omega)$ in
eqs. (\ref{fullP}) and (\ref{specdens}) by virtue of adiabaticity.
This explains that the experimental data show features at frequencies
close to the vibration-rotation excitations of the water molecule
\cite{priv}. Especially, one can expect this to be the dominant effect
in the spectra of multi-bubble sonoluminescence, as the resonances
discussed under {\em (i)} above will average out if bubbles grow and
collapse in a random manner.

\subsubsection{Predicted pulse length}
Apart from predicting the radiated spectrum, the theory of quantum
vacuum radiation solves several conceptual problems that previous
theories have not been able to deal with satisfactorily. Most
importantly, the present theory has no difficulty in explaining the
extreme shortness of the emitted light pulses; their duration is
determined by the parameter $\gamma$ describing the turn-around time
at the collapse and the time it takes for the fluctuations to
correlate around the bubble. As the latter is on the scale of merely
one or a few femtoseconds, $\gamma$ is the decisive quantity. Thus one
expects the pulse length to lie between 100fs and 10ps, which tallies
with the experimental observations.

\subsubsection{Absence of radiation in the UV}
Another major question that is successfully answered by the present theory
concerns the absence of radiation below the absorption edge of water
at around 180nm. Water has essentially no polarizability below this
wavelength, so that the real part of its refractive index is very
close to 1. Hence the mechanism of exciting vacuum fluctuations into
real photon pairs is inoperative below 180nm; no radiation is emitted
and no radiation has to be re-absorbed. This explains the absence of
any macroscopically discernible effect on the water by the large amounts
of absorbed light predicted by theories of black-body radiation or
bremsstrahlung (cf. Sec. I A).

\subsection{Suggested experiments}
Thinking about experiments that distinguish the present from other
theories of sonoluminescence, one quickly comes up with two relatively
simple ones. One is to look for photons emitted in the X-ray
transparency window of water \cite{debt}; both the black-body and the
bremsstrahlung theories predict a perceptible amount of photons with
wavelengths of around 1\AA, whereas the present theory denies any
photon emission at such short wavelengths since the polarizability of
water is essentially zero for X-rays, i.e. $(n-1)\approx 0$.

The second presumably easily set up experiment is to force the bubble
into an elongated rather than spherical shape by using piezoelectric
transducers on two or all three axes and to examine the angular
distribution of the emitted light. For such a case the present theory,
unlike others, predicts an anisotropic intensity; the number of
photons radiated into a given direction is roughly proportional to the
cross-section of the bubble perpendicular to that direction. Thus, if
the bubble is spheroidal rather than spherical during the radiation
process, one expects anisotropy.

\subsection{Agenda and open questions}
The most important point still to be attended to is to extend the
numerical calculations of the spectral density (\ref{fullP}) down to
realistic minimum bubble sizes of one micron or less
\cite{progress}. This will allow one to make more precise statements
as to the turn-around time $\gamma$ required to produce the
experimentally observed number of photons (cf. Sec. A {\em 1}) and to
explore the effects of resonances if the bubble size is comparable to
the photon wavelengths (cf. Sec. A {\em 3}).

Another effect to be studied in detail is the photon radiation
produced by the rapid variation of the refractive index of water due
to the rapidly varying compression around the outside of the bubble
\cite{progress}. Preliminary estimations have shown that this
mechanism is of secondary importance for the sonoluminescence problem;
to understand the principle of it might, however, be useful in
view of other applications.

An academic, but nevertheless interesting question to ask is where the
photons actually are produced. A model calculation for a
one-dimensional moving dielectric filling a half-space \cite{GBCE}
has indicated that, although the emission of this kind of quantum vacuum
radiation depends on the existence of a moving interface between two
media of different refractive indices, the photons do not come
directly from the interface but from within a certain vicinity of it,
as suggested also by physical intuition. However, inconsistencies
between where the photons are produced and which the support of the
radiation pressure on the dielectric is, are buried in the assumption
of a perfectly rigid dielectric.

The hydrodynamics of the bubble has been considered as given in the
present work; its theory has been quite successfully established
\cite{hydro}. However, especially in view of relaxing restrictions on
the velocity and the acceleration of the bubble surface in that theory
and of scrutinizing the bubble dynamics at the moment of collapse, the
important role of the back-reaction of the photon radiation onto the
bubble should be recognized. The momentum loss due to the radiation
process might have a significant effect on the bubble. While the
emission lasts the equations motion of the liquid-gas interface
will be supplemented by a frictional force which is roughly
proportional to the fourth derivative of the velocity of the
interface, as discussed just below eq. (\ref{Wres}).

One of the puzzles that remain is why stable single-bubble
sonoluminescence is seen only in water, although multi-bubble
sonoluminescence has been observed in a variety of fluids. The present
author's conjecture is that the reason for this is buried in the
unusual properties of gas solubility of water, which conspire with
hydrodynamic mechanisms to lead to a exceptionally sharp and violent
collapse of a driven bubble. In other fluids such conditions might be
reached at random, but not in a regular fashion to produce radiation
from stably maintained bubbles.

\subsection{Credo}
To close a conceptual remark might be appropriate. At first sight, the
idea that the burst of photons seen in sonoluminescence has its origin
in the zero-point fluctuations of the electromagnetic field might seem
utterly strange, as one tends to think of low-energy photons emitted
from material media as coming from atomic transitions. Pondering on
this, one has to admit that all we really know is that photons come
from some kind of moving charge or, field-theoretically speaking, from
the coupling to a fermion field. As we structure our thinking, we are
most inclined to consider atoms as the basic entities of all materials
and try to explain all physical phenomena on this basis. However,
there is no reason for so doing; we are completely at liberty to
mentally re-group these charges in a variety of different ways and
should choose whichever is most appropriate for the problem at hand.
In the case of sonoluminescence atoms are obviously not the basic
entities to be considered, since atomic transitions are about a
thousand times slower than a sonoluminescence pulse. Here the basic
structure of the medium with respect to the radiation process is most
suitably thought of as an assembly of dipoles with a certain
dielectric response. This point of view enables one to consider the
cooperative response of the charges to the zero-point fluctuations of
the electromagnetic field, and quantum vacuum radiation emerges as a
consequence quite naturally.

\section*{ACKNOWLEDGMENTS}
It is a pleasure to thank Peter W. Milonni for drawing my attention to
the phenomenon of sonoluminescence, for pointing out to me Julian
Schwinger's thought that this might relate to the zero-point
fluctuations of the photon field, and for urging me to apply my
knowledge from the theory of moving dielectrics to this
problem. Furthermore, I am grateful to Peter L. Knight for telling me
about the apparent thermal properties of pure two-mode states when
traced over one of the photons, and I would like to thank Gabriel
Barton, Nigel D. Goldenfeld, Anthony J. Leggett, Efrat Shimshoni, and
Shivaji L. Sondhi for their interest, stimulating discussions, and
many questions. I have also benefited from a discussion with Paul
M. Goldbart about the completeness of spherical solutions of the wave
equation. Financial support through the John D. and Catherine
T. MacArthur Foundation is gratefully acknowledged.

\appendix
%
\section*{APPENDIX A: HAMILTONIAN FOR A DIELECTRIC IN UNIFORM  MOTION}
\renewcommand{\theequation}{A\arabic{equation}}
\setcounter{equation}{0}
The standard way of deriving a Hamiltonian is to proceed from a
Lagrangian density. The Lagrangian density for a homogeneous
dielectric moving rigidly and uniformly is a function of the
dielectric constant $\varepsilon$ of the medium and of the velocity
$\beta$ of the medium relative to the frame of the observer;
\[ {\cal L}\;=\;{\cal L}(\varepsilon, \beta)\;. \]
It is uniquely determined by the following three requirements:

{\em (i)} In the limit of $\beta=0$ it should reduce to the familiar
Lagrangian density for a stationary dielectric
\begin{equation}
	{\cal L}(\varepsilon, \beta\!=\!0) = \frac{1}{2} \left(
	\frac{{\bf D}^2}{\varepsilon} - {\bf B}^2 \right)\;.
\label{requ1}
\end{equation}

{\em (ii)} For an optically transparent medium one should recover the
Lagrangian density of the vacuum, which, by Lorentz invariance, is
independent of the velocity $\beta$;
\begin{equation}
	{\cal L}(\varepsilon\!=\!1, \beta) = - \frac{1}{4}\:
	F_{\mu\nu} F^{\mu\nu}
	= \frac{1}{2} \left( {\bf E}^2 - {\bf B}^2 \right)\;.
\label{requ2}
\end{equation}
The symbol $F_{\mu\nu}$ denotes the field strength tensor of the
electromagnetic field, $F_{\mu\nu}=\nabla_{\mu}A_{\nu} -
\nabla_{\nu}A_{\mu}$. Its dual is defined $\tilde{F}_{\mu\nu} =
1/2\ \epsilon_{\mu\nu\alpha\beta}F^{\alpha\beta}$.

{\em (iii)} $\cal L$ must be a Lorentz scalar. The only true scalars
that are quadratic in the fields and that depend solely on the field
strength $F_{\mu\nu}$ and on the four-velocity $u_{\mu}$ of the medium
are $F_{\mu\nu} F^{\mu\nu}$, $u_{\mu} F^{\mu\nu} u^{\alpha}
F_{\alpha\nu}$, and $u_{\mu} \tilde{F}^{\mu\nu} u^{\alpha}
\tilde{F}_{\alpha\nu}$.

{}From the above the Lagrangian density is found to be
\begin{equation}
	{\cal L}(\varepsilon, \beta) = - \frac{1}{4}\:
	F_{\mu\nu} F^{\mu\nu} - \frac{\varepsilon - 1}{2}\;
	u_{\mu} F^{\mu\nu} u^{\alpha} F_{\alpha\nu}\;.
\label{lagr}
\end{equation}
Now the Hamiltonian can be derived by going through the canonical
formalism
\renewcommand{\theequation}{A\arabic{equation}\alph{multieqn}}%
\setcounter{multieqn}{1}%
\begin{eqnarray}
	{\bf \Pi} &=& \frac{\delta{\cal L}}{\delta \dot{\bf A}}\;,
\label{canmom}\\
\addtocounter{equation}{-1} \addtocounter{multieqn}{1}%
	{\cal H} &=& {\bf \Pi} \cdot \dot{\bf A} - {\cal L}\;.
\label{canham}
\end{eqnarray}\renewcommand{\theequation}{A\arabic{equation}}
This leads to the Hamiltonian density
\begin{equation}
	{\cal H} = \frac{1}{2}\;
	\frac{\varepsilon(1-\beta^2)}{\varepsilon - \beta^2} \left(
	\frac{{\bf \Pi}^2}{\varepsilon} + {\bf B}^2 \right)
	- \frac{\varepsilon-1}{\varepsilon - \beta^2}\:
	{\bf \beta} \cdot ({\bf \Pi}\wedge{\bf B})
	+ \frac{1}{2}\;
	\frac{\varepsilon-1}{\varepsilon - \beta^2} \left[
	\frac{({\bf \beta}\cdot{\bf \Pi})^2}{\varepsilon} +
	({\bf \beta}\cdot{\bf B})^2 \right]\:.
\label{hdens}
\end{equation}
Substituting ${\bf \Pi}=-{\bf D}$ and, with a perturbative treatment
in mind, expanding in powers of the velocity $\beta$, one obtains
\begin{equation}
	H = \int{\rm d}^3{\bf r} \left[ \frac{1}{2} \left(
	\frac{{\bf D}^2}{\varepsilon} + {\bf B}^2 \right)
	+ \frac{\varepsilon -1}{\varepsilon}\:
	{\bf \beta} \cdot ({\bf D}\wedge{\bf B}) + {\rm O}(\beta^2)
	\right]\;.
\label{ham}
\end{equation}
This is a very natural result; the Hamiltonian for a stationary
dielectric is augmented by an energy-fluxlike correction which
vanishes for a transparent medium.

Another way of arriving at the same result is to appeal to the
Lorentz invariance of the Maxwell theory. The Hamiltonian density must
in any frame be given by
\begin{equation}
	{\cal H} = \frac{1}{2} \left( {\bf D}\cdot{\bf E} +
	{\bf B}\cdot{\bf H} \right)\;,
\label{anyframe}
\end{equation}
where the fields are as measured in this frame. For a non-magnetic
dielectric ${\bf B}={\bf H}$, but the ${\bf D}$ and ${\bf E}$ fields
are connected by some non-trivial constitutive relation which can be
found by Lorentz-transforming the constitutive relation ${\bf D}' =
\varepsilon{\bf E}'$ valid in the rest-frame of the medium into the
laboratory frame. There the constitutive relations read
\renewcommand{\theequation}{A\arabic{equation}\alph{multieqn}}%
\setcounter{multieqn}{1}%
\begin{eqnarray}
	{\bf E}_{\parallel}&=& \frac{1}{\varepsilon}\:{\bf D}_{\parallel}\;,
\label{const1}\\
\addtocounter{equation}{-1} \addtocounter{multieqn}{1}%
	{\bf E}_{\perp}&=& \frac{\varepsilon}{\varepsilon-\beta^2}\:
	\left[ \frac{1}{\varepsilon}\:(1-\beta^2)\:{\bf D}_{\perp}
	- \frac{\varepsilon-1}{\varepsilon}\:{\bf \beta}\wedge{\bf B}
	\right] \;.
\label{const2}
\end{eqnarray}\renewcommand{\theequation}{A\arabic{equation}}
Utilizing these to replace {\bf E} and {\bf H} in eq (\ref{anyframe})
one recovers the Hamiltonian density (\ref{hdens}) obtained earlier
by different means.

\section*{APPENDIX\hspace{2mm} B\hspace{1mm}:\hspace{2mm} MODE\hspace{2mm}
EXPANSION\hspace{2mm} FOR\hspace{2mm} THE HELMHOLTZ EQUATION IN
SPHERICAL COORDINATES}
\renewcommand{\theequation}{B\arabic{equation}}
\setcounter{equation}{0}
The mode functions $A_{(1,2)}^{\rm TE,TM}$ in the expansions (\ref{mode})
are solutions of the Helmholtz equation
\begin{equation}
	\omega^2\,\varepsilon({\bf r})\,{\bf A}({\bf r},{\bf k})
	+ \nabla^2\,{\bf A}({\bf r},{\bf k}) = 0 \;,\;\;
	\omega = | {\bf k} | \;.
\label{helm}
\end{equation}
Rewriting this equation as
\begin{equation}
	\left( \frac{1}{\sqrt{\varepsilon}} \nabla^2
	\frac{1}{\sqrt{\varepsilon}} \right)\,\sqrt{\varepsilon}{\bf A}
	= -\omega^2\,\sqrt{\varepsilon}{\bf A}
\label{herm}
\end{equation}
makes obvious that this is the eigenvalue equation of the Hermitean
operator $(1/\sqrt{\varepsilon}) \nabla^2 (1/\sqrt{\varepsilon})$, and
hence the mode functions $\sqrt{\varepsilon}{\bf A}$ form a complete
set of orthogonal functions.

In order to diagonalize the Hamiltonian (\ref{h0}) into the Hamiltonian
(\ref{hphot}) of the photon field by means of the mode expansions
(\ref{mode}), the mode functions should satisfy the orthonormalization
conditions
\begin{equation}
	\int{\rm d}^3 {\bf r}\; \varepsilon({\bf r}) \left[
	{\bf A}_{(1)}^{\rm TE,TM}({\bf r};{\bf k})
	{\bf A}_{(1)}^{{\rm TE,TM}\ast}({\bf r};{\bf k'}) +
	{\bf A}_{(2)}^{\rm TE,TM}({\bf r};{\bf k})
	{\bf A}_{(2)}^{{\rm TE,TM}\ast}({\bf r};{\bf k'}) \right]
	= \delta^{(3)}({\bf k}-{\bf k'})
\label{norm}
\end{equation}
for each the TE and the TM polarizations.

To find the solutions of the Helmholtz equation (\ref{helm}) one
conveniently proceeds from the scalar solution
\begin{equation}
   \Phi = \frac{4\pi}{(2\pi)^{3/2}} \sum_{\ell,m}
	  {\rm e}^{-{\rm i}\delta_{\ell}}\;
          {\rm i}^{\ell} \left[ \cos \delta_{\ell}\; j_{\ell}(nkr) + \sin
          \delta_{\ell}\; y_{\ell}(nkr) \right] Y_{\ell}^{m\ast}(\hat{\bf k})
          Y_{\ell}^{m}(\hat{\bf r})
\label{scalar}
\end{equation}
which is normalized to behave like a plane wave
${\rm e}^{{\rm i}n{\bf k}\cdot{\bf r}}/(2\pi)^{3/2}$
for $kr \rightarrow \infty$. The phase $\delta_{\ell}$ will be
chosen to meet the required continuity conditions across the surface of the
bubble.

The vector field operators invariant under rotation are ${\bf r}$,
${\bf \nabla}$, ${\bf L}=-{\rm i}{\bf r}\times{\bf \nabla}$, and ${\bf
\nabla}\times{\bf L}$. Since ${\bf r}$ fails to commute with
$\nabla^{2}$, only the last three operators may be used to generate
rotation-invariant vector solutions of the Helmholtz equation.
However, ${\bf \nabla}\Phi$ is an irrotational field; only the
solenoidal fields ${\bf L}\Phi$ and $({\bf \nabla}\times{\bf L})\Phi$
can be employed for representations of the transversely polarized
electromagnetic field. Choosing ${\bf A}_{(1)} \sim {\bf L}\Phi$ and
${\bf A}_{(2)} \sim 1/(nk)\; {\bf \nabla}\times{\bf L} \Phi$ and
observing the correct normalization (\ref{norm}) one obtains for the
mode functions outside the bubble
\begin{equation}
   \begin{array}{l}
      {\displaystyle
   A_{(1)\theta}=\sqrt{\frac{n}{\pi}}\; \sum_{\ell,m}\;
\frac{{\rm e}^{-{\rm i}\delta_{\ell}}\;
   {\rm i}^{\ell}}{\sqrt{\ell(\ell+1)}} \;\left[ \cos \delta_{\ell}\;
j_{\ell}(nkr)
   + \sin \delta_{\ell}\; y_{\ell}(nkr) \right]\;
Y_{\ell}^{m\ast}(\hat{\bf k})\;
   \frac{(-m)}{\sin\theta}\; Y_{\ell}^{m}(\hat{\bf r}) }\;,\\
      {\displaystyle
   A_{(1)\varphi}=\sqrt{\frac{n}{\pi}}\; \sum_{\ell,m}\;
\frac{{\rm e}^{-{\rm i}\delta_{\ell}}\;
   {\rm i}^{\ell}}{\sqrt{\ell(\ell+1)}} \;\left[ \cos \delta_{\ell}\;
j_{\ell}(nkr)
   + \sin \delta_{\ell}\; y_{\ell}(nkr) \right]\;
Y_{\ell}^{m\ast}(\hat{\bf k})\;
   (-{\rm i})\; \frac{\partial Y_{\ell}^{m}(\hat{\bf r})}{\partial\theta}
}\;,\\
   A_{(1) r}=0 \;,\\
      {\displaystyle
   A_{(2)\theta}=\sqrt{\frac{n}{\pi}}\; \sum_{\ell,m}\;
\frac{{\rm e}^{-{\rm i}\delta_{\ell}}\;
   {\rm i}^{\ell}}{\sqrt{\ell(\ell+1)}}\; \frac{1}{nkr} \;\left[ nkr\:\cos
\delta_{\ell}
   \;j_{\ell}(nkr) + nkr\:\sin \delta_{\ell}\; y_{\ell}(nkr)
\right]^{\prime} }\\
      {\displaystyle
   \hspace*{110mm} \times Y_{\ell}^{m\ast}(\hat{\bf k})\;{\rm i}\;
   \frac{\partial Y_{\ell}^{m}(\hat{\bf r})}{\partial\theta} }\;,\\
      {\displaystyle
   A_{(2)\varphi}=\sqrt{\frac{n}{\pi}}\; \sum_{\ell,m}\;
\frac{{\rm e}^{-{\rm i}\delta_{\ell}}\;
   {\rm i}^{\ell}}{\sqrt{\ell(\ell+1)}} \;\frac{1}{nkr} \;\left[ nkr\:\cos
\delta_{\ell}
   \;j_{\ell}(nkr) + nkr\:\sin \delta_{\ell}\; y_{\ell}(nkr)
\right]^{\prime} }\\
      {\displaystyle
   \hspace*{110mm} \times Y_{\ell}^{m\ast}(\hat{\bf k})\;
   \frac{(-m)}{\sin\theta}\;Y_{\ell}^{m}(\hat{\bf r}) }\;,\\
      {\displaystyle
   A_{(2) r}=\sqrt{\frac{n}{\pi}}\; \sum_{\ell,m}\;
\frac{{\rm e}^{-{\rm i}\delta_{\ell}}\;
   {\rm i}^{\ell}}{\sqrt{\ell(\ell+1)}} \frac{{\rm i}\ell(\ell+1)}{nkr}
   \;\left[ \cos \delta_{\ell}\;j_{\ell}(nkr) + \sin \delta_{\ell}
   \; y_{\ell}(nkr) \right]\; Y_{\ell}^{m\ast}(\hat{\bf k})\:
Y_{\ell}^{m}(\hat{\bf r}) }\;,
   \end{array}
\label{outs}
\end{equation}
and for those inside the bubble
\begin{equation}
   \begin{array}{l}
      {\displaystyle
   A_{(1)\theta}=\sqrt{\frac{n}{\pi}}\; \sum_{\ell,m}\; {\cal S}^{-1}_{\ell}\;
   \frac{{\rm i}^{\ell}}{\sqrt{\ell(\ell+1)}} \;j_{\ell}(kr)
   \; Y_{\ell}^{m\ast}(\hat{\bf k})\;
   \frac{(-m)}{\sin\theta}\; Y_{\ell}^{m}(\hat{\bf r}) }\;,\\
      {\displaystyle
   A_{(1)\varphi}=\sqrt{\frac{n}{\pi}}\; \sum_{\ell,m}\;{\cal S}^{-1}_{\ell}\;
   \frac{{\rm i}^{\ell}}{\sqrt{\ell(\ell+1)}} \;j_{\ell}(kr)
   \; Y_{\ell}^{m\ast}(\hat{\bf k})\;
   (-{\rm i})\; \frac{\partial Y_{\ell}^{m}(\hat{\bf r})}{\partial\theta}
}\;,\\
   A_{(1) r}=0 \;,\\
      {\displaystyle
   A_{(2)\theta}=\sqrt{\frac{n}{\pi}}\; \sum_{\ell,m}\; {\cal S}^{-1}_{\ell}\;
   \frac{{\rm i}^{\ell}}{\sqrt{\ell(\ell+1)}}\; \frac{1}{kr} \;\left[ kr
   \;j_{\ell}(kr) \right]^{\prime} \;Y_{\ell}^{m\ast}(\hat{\bf k})\;{\rm i}\;
   \frac{\partial Y_{\ell}^{m}(\hat{\bf r})}{\partial\theta} }\;,\\
      {\displaystyle
   A_{(2)\varphi}=\sqrt{\frac{n}{\pi}}\; \sum_{\ell,m}\; {\cal S}^{-1}_{\ell}\;
   \frac{{\rm i}^{\ell}}{\sqrt{\ell(\ell+1)}} \;\frac{1}{kr} \;\left[ kr
   \;j_{\ell}(kr) \right]^{\prime} \;Y_{\ell}^{m\ast}(\hat{\bf k})\;
   \frac{(-m)}{\sin\theta}\;Y_{\ell}^{m}(\hat{\bf r}) }\;,\\
      {\displaystyle
   A_{(2) r}=\sqrt{\frac{n}{\pi}}\; \sum_{\ell,m}\; {\cal S}^{-1}_{\ell}\;
   \frac{{\rm i}^{\ell}}{\sqrt{\ell(\ell+1)}} \frac{{\rm i}\ell(\ell+1)}{kr}
   \; j_{\ell}(kr)
   \; Y_{\ell}^{m\ast}(\hat{\bf k})\:Y_{\ell}^{m}(\hat{\bf r}) }\;.
   \end{array}
\label{ins}
\end{equation}
Here and in the following a prime behind a bracket means a derivative
with respect to the argument of the spherical Bessel function.
The mode functions for the inside of the bubble have zero phase shift
because the fields have to be regular at the origin $r=0$, which
excludes any contributions from the spherical Bessel functions of the
second kind $y_{\ell}$ as these diverge for zero argument.

The phase shifts $\delta_{\ell}^{\rm TE,TM}$ and the normalization
constants ${\cal S}^{{\rm TE,TM}\:-1}_{\ell}$ are determined by the
continuity conditions (\ref{contexpl}) across the bubble surface at $r=R$.
One obtains
\begin{equation}
	\tan \delta_{\ell}^{\rm TE} =
	\frac{{\cal N}_{\ell}^{\rm TE}}{{\cal D}_{\ell}^{\rm TE}}\;,
	\hspace*{5mm}
	\tan \delta_{\ell}^{\rm TM} =
	\frac{{\cal N}_{\ell}^{\rm TM}}{{\cal D}_{\ell}^{\rm TM}}\;,
\label{deltas}
\end{equation}
and
\begin{equation}
	{\cal S}^{{\rm TE}\;-1}_{\ell} = \frac{1}{(nkR)}\;
	\frac{1}{(-{\cal D}_{\ell}^{\rm TE}
	-{\rm i}\:{\cal N}_{\ell}^{\rm TE})}\;,
	\hspace*{5mm}
	{\cal S}^{{\rm TM}\;-1}_{\ell} = \frac{1}{(kR)}\;
	\frac{1}{(-{\cal D}_{\ell}^{\rm TM}
	-{\rm i}\:{\cal N}_{\ell}^{\rm TM})}\;,
\label{STEM}
\end{equation}
where
\begin{equation}
\begin{array}{l}
	{\cal N}_{\ell}^{\rm TE} =
	j_{\ell}(kR) \left[ nkR \;j_{\ell}(nkR) \right]^{\prime}
	- j_{\ell}(nkR) \left[ kR \;j_{\ell}(kR) \right]^{\prime}\;,\\
	{\cal D}_{\ell}^{\rm TE} =
	y_{\ell}(nkR) \left[ kR \;j_{\ell}(kR) \right]^{\prime}
	- j_{\ell}(kR) \left[ nkR \;y_{\ell}(nkR) \right]^{\prime}\;,\\
	{\cal N}_{\ell}^{\rm TM} =
	j_{\ell}(kR) \left[ nkR \;j_{\ell}(nkR) \right]^{\prime}
	- n^2\;j_{\ell}(nkR) \left[ kR \;j_{\ell}(kR) \right]^{\prime}\;,\\
	{\cal D}_{\ell}^{\rm TM} =
	n^2\;y_{\ell}(nkR) \left[ kR \;j_{\ell}(kR) \right]^{\prime}
	- j_{\ell}(kR) \left[ nkR \;y_{\ell}(nkR) \right]^{\prime}\;.
\end{array}
\label{NandD}
\end{equation}
Note that
\begin{equation}
	\lim_{n\rightarrow 1}\;{\cal S}^{{\rm TE,TM}\;-1}_{\ell} = 1\;,
\label{Born}
\end{equation}
due to the fact that in this limit the ${\cal D}_{\ell}^{\rm TE,TM}$
are simplified by the Wronskian of the spherical Bessel functions
\[ j_{\ell}(x) y_{\ell}^{\prime}(x) - j_{\ell}^{\prime}(x) y_{\ell}(x) =
     \frac{1}{x^{2}}\;, \]
and the ${\cal N}_{\ell}^{\rm TE,TM}$ obviously reduce to zero.

\section*{APPENDIX\hspace{2mm} C\hspace{1mm}:\hspace{2mm}
FORCE\hspace{2mm} ON\hspace{2mm} A\hspace{2mm} STATIONARY DIELECTRIC}
\renewcommand{\theequation}{C\arabic{equation}}
\setcounter{equation}{0}
There are several ways of deriving an expression for the force applied
by an electromagnetic field on a dielectric body; the physically most
intuitive one is to consider the force as ensuing from induced
currents and surface-charge densities.

In Lorentz gauge or in Coulomb gauge without free charges the vector
potential satisfies the wave equation
\begin{equation}
	\frac{\partial}{\partial t}\:(\varepsilon\dot{\bf A})
	- \nabla^2 {\bf A} = 0\;.
\label{wave}
\end{equation}
On rewriting this equation as
\[	\ddot{\bf A} - \nabla^2 {\bf A} = {\bf j}_{\rm ind}\;,  \]
one obtains an induced current density
\begin{equation}
	{\bf j}_{\rm ind} = - (\varepsilon -1) \ddot{\bf A}\;.
\label{jind}
\end{equation}
By continuity
\[ 	\frac{\partial\rho_{\rm ind}}{\partial t} + \nabla\cdot
   	{\bf j}_{\rm ind} = 0  \]
one finds that the induced surface-charge density is
\begin{equation}
	\rho_{\rm ind} = - \nabla\cdot \left( \frac{\varepsilon
	-1}{\varepsilon} {\bf D} \right)\;.
\label{rhoind}
\end{equation}
Therefore a stationary dielectric is acted upon by a force density
\begin{equation}
	{\bf f} = \rho_{\rm ind} {\bf E} + {\bf j}_{\rm ind}
	\wedge {\bf B}\;.
\label{forcedens}
\end{equation}
On integrating this density over a dielectric with a cavity one has to
bear in mind that the radial component of the electric field is not
continuous across the boundary. However, the gradient of
$(\varepsilon-1)/\varepsilon$ in (\ref{rhoind}) brings about a
$\delta$ function on the surface of the cavity, which is multiplied
by the electric field in eq (\ref{forcedens}) for the force
density. The mathematically and physically correct prescription is to
substitute the electric field by the average of its values on the two
sides of the boundary. Then one obtains for the radial component of the
force on a spherical bubble of radius $R$
\begin{equation}
	{\cal F}_{r} = - \left( 1-\frac{1}{n^2} \right)
	\frac{R^2}{2} \oint {\rm d}\Omega \left[ \left(
	1+\frac{1}{n^2}\right) D_r^2 + B_r^2 - B_\theta^2
	- B_\phi^2 \right]
	\;,
\label{force}
\end{equation}
where the integral runs over the complete solid angle; the tangential
components of the force are zero as obvious from symmetry.

Strictly speaking, the expression (\ref{force}) is of course the force
on the dielectric and not the force on the bubble. Nevertheless, since
the two are complementary and the force density has $\delta$-support
on the boundary, i.e. the force is really applied only to the
interface, it seems reasonable to speak of the force as acting on the
bubble.

Another, less palpable and more formal way of calculating the force
density (\ref{forcedens}) is to proceed from the
stress-energy-momentum tensor of the electromagnetic field. In this
approach care must be taken when interpreting formulae, because the
momentum density of the photon field in a dielectric medium is subject
to ambiguity. However, this issue will not be addressed here as the
force density is unequivocally defined and interpretable.

The stress exerted by the fields alone, i.e. the fields as in vacuum
and exclusive of the polarization fields inside the dielectric, is
given by the space-like components of the stress-energy-momentum
tensor in vacuum
\begin{equation}
	T_{(0)}^{ij}(\varepsilon\!=\!1) = -E_i E_j - B_i B_j
	+ \frac{1}{2}\,\delta_{ij}\, ({\bf E}^2 + {\bf B}^2)\;.
\label{Tvac}
\end{equation}
In vacuum there is no doubt about the momentum density carried by the
field; it reads
\begin{equation}
	T_{(0)}^{i0}(\varepsilon\!=\!1) = \epsilon_{ijk} E_j B_k\;,
\label{momvac}
\end{equation}
which is just the Poynting vector. Overall momentum balance requires
that the change in the mechanical momentum density of the material
(i.e. the force density on the dielectric) together with the change in
the momentum of the fields alone (\ref{momvac}) equal the negative
gradient of the stress (\ref{Tvac}) due to the fields,
\[ \frac{\partial}{\partial t} \left[ {\cal M}_{\rm mech}^i +
   T^{i0}_{(0)} (\varepsilon\! =\!1) \right] \, =\, -\nabla_j
   T^{ij}_{(0)} (\varepsilon\! =\!1)\;. \]
Thus the force density is given by
\[ 	f_i = \frac{\partial{\cal M}_{\rm mech}^i}{\partial t}
	= - \frac{\partial}{\partial t}T^{i0}_{(0)}
	-\nabla_j T^{ij}_{(0)} \;. \]
A few trivial transformations taking into account vector identities
and the Maxwell equations yield
\[ 	f_i = \left( 1-\frac{1}{\varepsilon}\right) \left( B_k
	\nabla_k B_i - \frac{1}{2}\, \nabla_i {\bf B}^2 \right) + E_i
	\nabla_j E_j \;. \]
Upon integration over the bubble the first part of this expression is
easily seen to lead to the $B$-dependent terms in the force
(\ref{force}). In order to recognize the second part one should note that
$\nabla_j E_j \equiv \nabla \cdot {\bf E}\ $ is zero inside as well as
outside the dielectric, but non-zero at the interface; $\nabla_j E_j$
gives a $\delta$ function on the surface multiplied by the difference
of the outer and the inner electric fields. Thus one recovers the induced
surface-charge density (\ref{rhoind}), the force on which engenders
the same $D_r$-dependent term as in (\ref{force}) before.


\newpage
\renewcommand{\baselinestretch}{2.0}
\section*{Figure captions}
\subsection*{Figure 1}
The model profile of eq. (4.9) for the squared radius
at the collapse of the bubble, printed for four different values of
the parameter $\gamma$. The solid line corresponds to the function
with the largest $\gamma$.
\subsection*{Figure 2}
The spectral density calculated numerically from eq. (4.5)
as a function of photon frequency, for the model profile (4.9) with
$R_0=45\mu$m, $R_{\rm min}=3\mu$m, and $\gamma=10$fs.
\subsection*{Figure 3}
The same data as in figure 2 but as a function of photon
wavelength. Note that one can barely make out the features that are
clearly visible in figure 2.

\begin{thebibliography}{99}
\bibitem{first} H. Frenzel and H. Schultes, Z. Phys. Chem., Abt. B
	{\bf 27}, 421 (1934).
\bibitem{stable1} D. F. Gaitan, L. A. Crum, C. C. Church, and R. A.
	Roy, J. Acoust. Soc. Am. {\bf 91}, 3166 (1992).
\bibitem{stable2} B. P. Barber and S. J. Putterman, Nature {\bf 352},
	318 (1991); B. P. Barber, R. Hiller, K. Arisaka, H. Fetterman,
	and S. J. Putterman, J. Acoust. Soc. Am. {\bf 91}, 3061 (1992).
\bibitem{chaos} R. G. Holt, D. F. Gaitan, A. A. Atchley, and J.
	Holzfuss, Phys. Rev. Lett. {\bf 72}, 1376 (1994).
\bibitem{spec} R. Hiller, S. J. Putterman, and B. P. Barber, Phys.
	Rev. Lett. {\bf 69}, 1182 (1992).
\bibitem{hydro} R. L\"{o}fstedt, B. P. Barber, and S. J. Putterman,
	Phys. Fluids A {\bf 5}, 2911 (1993).
\bibitem{miemeas} B. P. Barber and S. J. Putterman, Phys. Rev. Lett.
	{\bf 69}, 3839 (1992).
\bibitem{brems} C. C. Wu and P. H. Roberts, Proc. R. Soc. Lond. A {\bf
	445}, 323 (1994).
\bibitem{priv} K. S. Suslick (private communication).
\bibitem{susl} E. B. Flint and K. S. Suslick, Science {\bf 253}, 1397
	(1991); and references therein.
\bibitem{fareast} Y. T. Didenko, T. V. Gordeychuk, and V. L. Koretz,
        J. Sound Vibr. {\bf 147}, 409 (1991).
\bibitem{collis} L. Frommhold and A. A. Atchley, Phys. Rev. Lett. {\bf
        73}, 2883 (1994).
\bibitem{jschw} J. Schwinger, Proc. Natl. Acad. Sci. USA {\bf 89},
        4091 (1992); {\bf 89}, 11118 (1992); {\bf 90}, 958 (1993);
	{\bf 90}, 2105 (1993); {\bf 90}, 4505 (1993).
\bibitem{casi} H. B. G. Casimir, Proc. K. Ned. Akad. Wet. {\bf 51},
        793 (1948).
\bibitem{milo} P. W. Milonni, {\em The Quantum Vacuum} (Academic, New
        York, 1994).
\bibitem{unruh} W. G. Unruh, Phys. Rev. {\bf D 14}, 870 (1976).
\bibitem{birdav} N. D. Birrell and P. C. W. Davies, {\em Quantum
        Fields in Curved Space} (Cambridge University Press, 1982).
\bibitem{GBCE} G. Barton and C. Eberlein, Ann. Phys. (N.Y.) {\bf 227},
        222 (1993).
\bibitem{fn1} In this way mean-square deviations of Casimir forces have
	been calculated for an infinite plane mirror by G. Barton,
	J. Phys. A: Math. Gen. {\bf 24}, 991 (1991); and for mirrors
	of finite size such as spheres and circular disks by C.
	Eberlein \cite{CEdf2}.
\bibitem{CEfluc} C. Eberlein, J. Phys. I (France) {\bf 3}, 2151 (1993).
\bibitem{knight} S. M. Barnett and P. L. Knight, J. Opt. Soc. Am. B
	{\bf 2}, 467 (1985); Phys. Rev. {\bf A 38}, 1657 (1988).
\bibitem{umez} H. Umezawa, H. Matsumoto, and M. Tachiki, {\em Thermo
	Field Dynamics and Condensed States} (North-Holland,
	Amsterdam, 1982); H. Umezawa, {\em Advanced Field Theory :
	Micro, Macro, and Thermal physics} (Am. Inst. Phys., New York, 1993).
\bibitem{yabl} E. Yablonovitch, Phys. Rev. Lett. {\bf 62}, 1742 (1989).
\bibitem{thesis} C. Eberlein, {\em Zero-Point Fluctuations and Quantum
	Radiation by Moving Mirrors}, D.~Phil. thesis (University of
	Sussex, 1993).
\bibitem{GR} I. S. Gradshteyn and I. M. Ryshik, {\em Tables of
	Integrals, Series, and Products} (Academic, New York, 1980).
\bibitem{AS} M. Abramowitz and I. A. Stegun (eds.), {\em Handbook of
	Mathematical Functions} (US Govt Printing Office, Washington
	DC, 1964).
\bibitem{pauli} W. Pauli, {\em Die allgemeinen Prinzipien der Wellenmechanik},
      	in ``Handbuch der Physik'', vol. V.1, ed. S. Fl\"{u}gge (Springer,
	Berlin, 1958), section 11.
\bibitem{schiff} L. I. Schiff, {\em Quantum Mechanics}, 3rd ed.
	(McGraw-Hill, Singapore, 1968), eq. (35.26).
\bibitem{fn2} Note especially that for $|m\rangle = |k,k'; R\rangle$
	the second term on the left-hand side of (\ref{key}) is of
	order $\beta^{2}$ and can therefore be neglected. Otherwise it
	would be awkward to deal with, since the argument normally
	used for non-degenerate states, that $\langle m|(\partial/
	\partial R)|m\rangle$ is purely imaginary and can thus be
	absorbed into a physically non-significant phase, does not
	apply in this case.
\bibitem{paulo} P. A. Maia Neto, J. Phys. A: Math. Gen. {\bf 27}, 2167 (1994).
\bibitem{progress} C. Eberlein (work in progress).
\bibitem{noble} R. Hiller, K. Weninger, S. J. Putterman, and B. P. Barber,
	Science {\bf 266}, 248 (1994).
\bibitem{CEdf2} C. Eberlein, J. Phys. A: Math. Gen. {\bf 25} 3015 (1992);
	{\bf 25} 3039 (1992).
\bibitem{debt} The author is indebted to Ken S. Suslick for pointing this out.
\end{thebibliography}
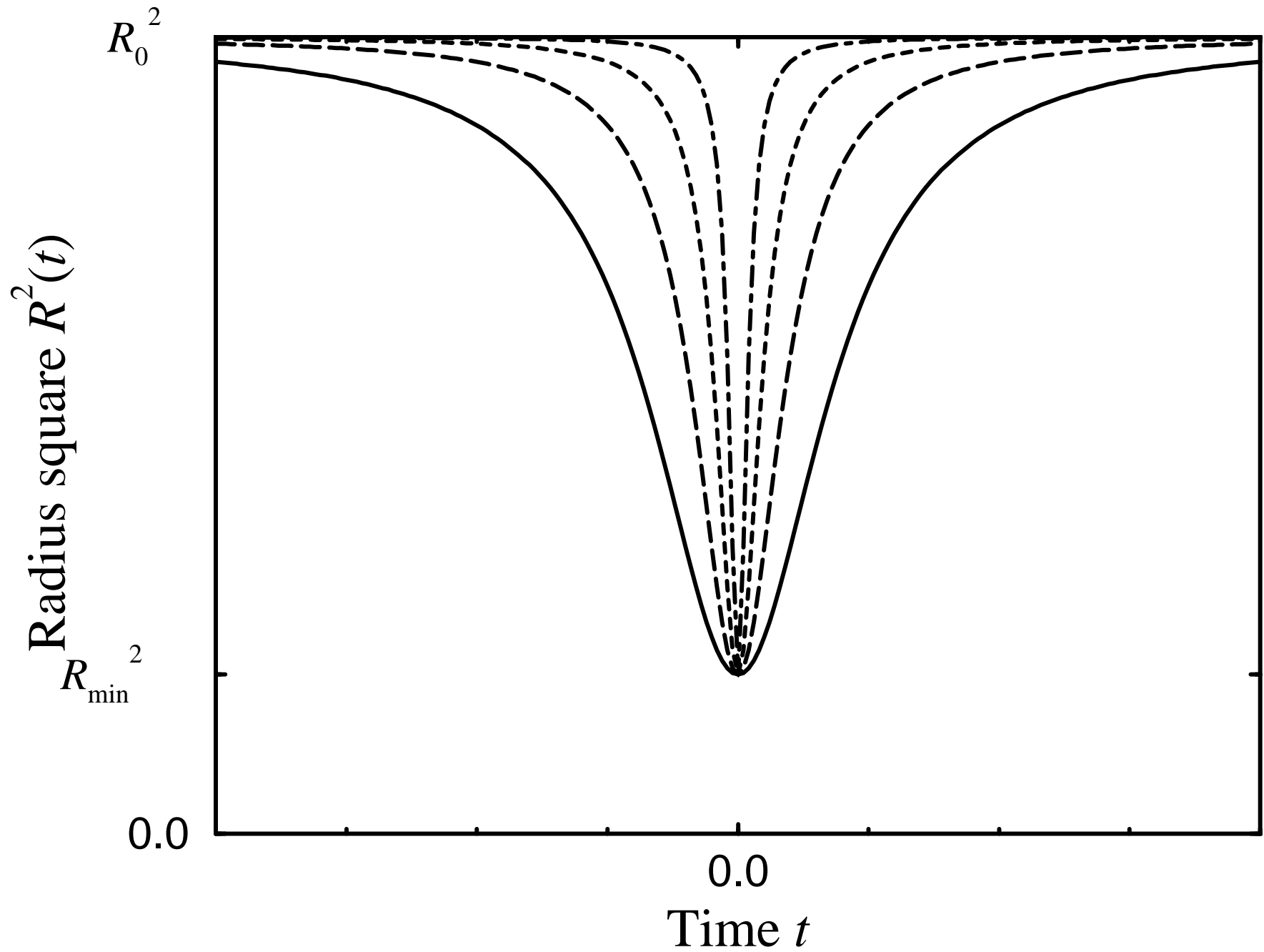
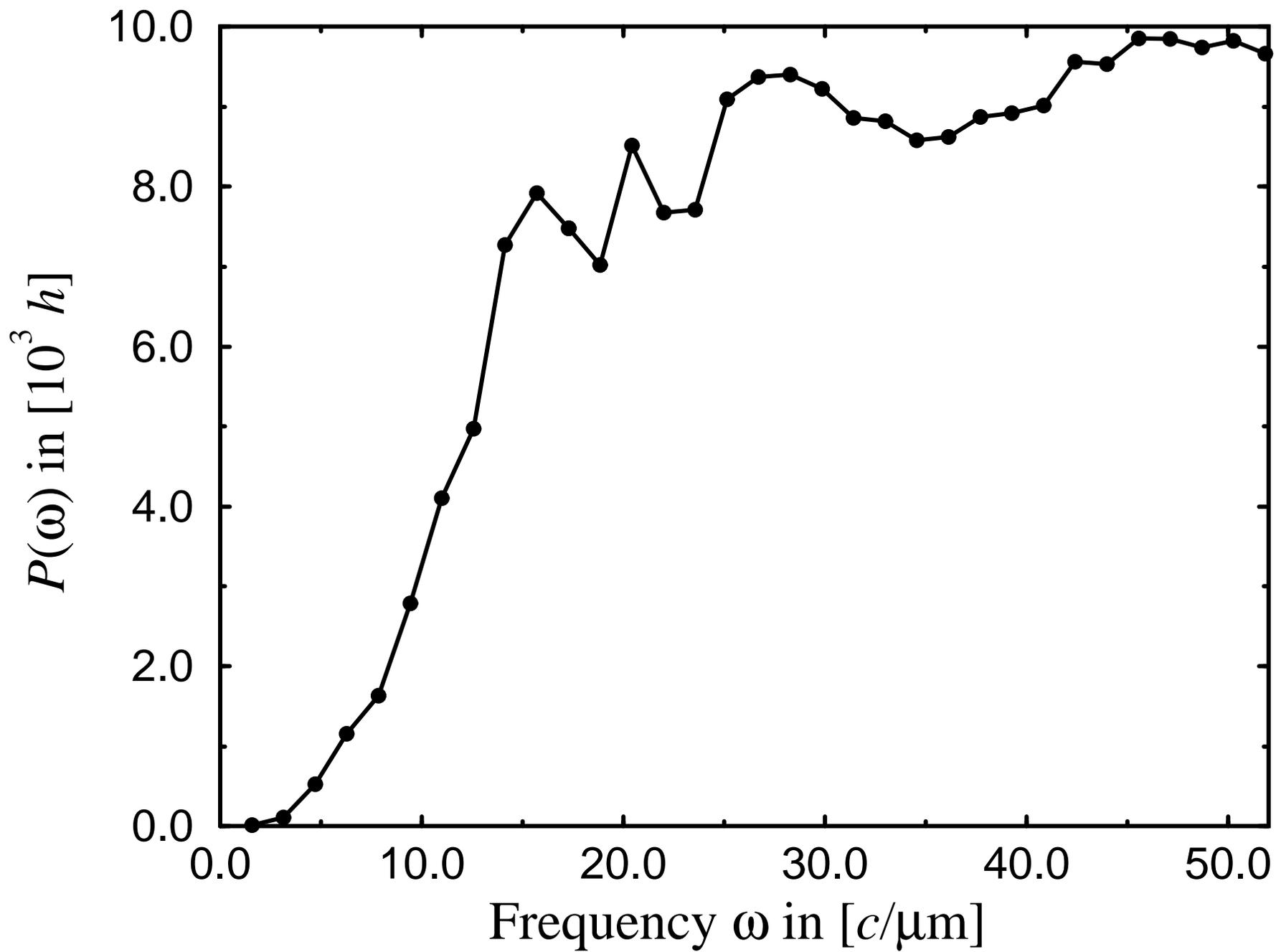
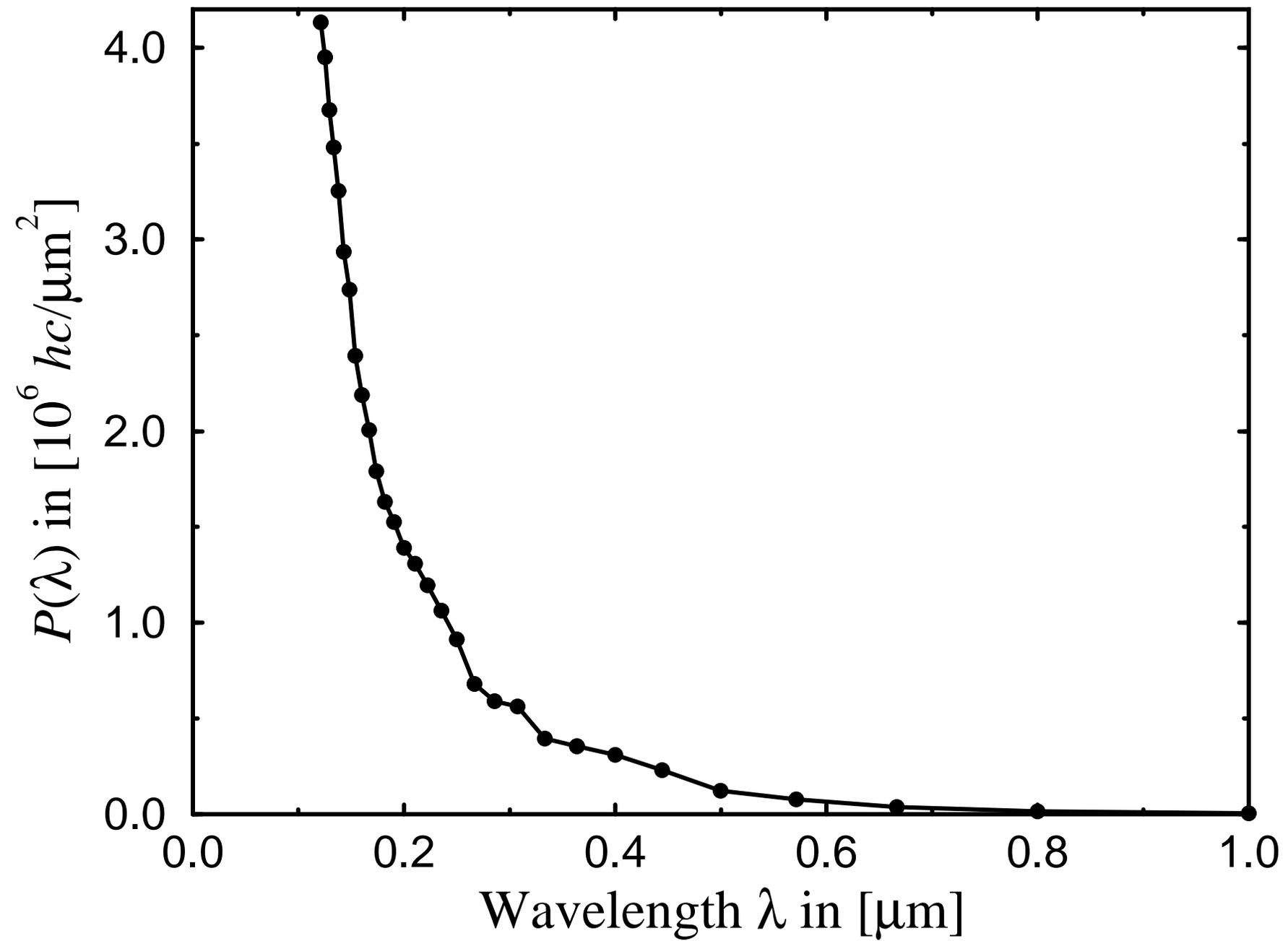
\end{document}